\documentclass[10pt]{iopart}

\expandafter\let\csname equation*\endcsname\relax
\expandafter\let\csname endequation*\endcsname\relax
\usepackage{amsmath}
\usepackage{iopams}
\usepackage{graphicx}
\usepackage{mathtools, cuted}
\usepackage{cleveref}
\usepackage[lofdepth,lotdepth]{subfig}
\usepackage{blindtext}
\usepackage{color}
\usepackage{cite}
\usepackage{lipsum}
\usepackage{breqn}
\usepackage{url}

\newcommand{\R}[1]{{#1}}

\newcommand{\mb}[1]{\mbox{\boldmath $#1$}}
\newcommand{\vE}{\mb{E}}
\newcommand{\vJ}{\mb{J}}
\newcommand{\vB}{\mb{B}}
\newcommand{\vA}{\mb{A}}
\newcommand{\vDJ}{\Delta\mb{J}}
\newcommand{\Dt}{\Delta t}
\newcommand{\dif}{{\rm d}}
\newcommand{\dvol}{{\rm d}^3\mb{r}}

\begin{document}

\title {Modelling the mechanics of 32 T REBCO superconductor magnet using numerical simulation}

\author{Arpit Kumar Srivastava \& Enric Pardo}

\address{Institute of Electrical Engineering, Slovak Academy of Sciences, Dubravka 9,84104 Bratislava, Slovakia\\
E-mail: arpit.srivastava@savba.sk \& enric.pardo@savba.sk\\
\vspace{3mm}
This is the Accepted Manuscript version of an article accepted for publication in Superconductor Science and Technology. IOP Publishing Ltd is not responsible for any errors or omissions in this version of the manuscript or any version derived from it. The Version of Record is available online at \url{https://doi.org/10.1088/1361-6668/ad4a34}. This manuscript is hereby available under license CC BY-NC-ND.}
\vspace{10pt}


\begin{abstract}
High temperature REBCO superconducting tapes are very promising for high-field magnets. With high magnetic field application there are high electro-mechanical forces, and thus concern for mechanical damage. Due to the presence of large screening currents and composite structure of the tape, the mechanical design of  these magnets are not straight forward. In addition, many contemporary designs use insulated winding. In this work we develop a novel two-dimensional axisymmetric finite element tool programmed in MATLAB that assumes the displacement field within linear elastic range. The stack of pancakes and a large number of REBCO tape turns are approximated as an an-isotropic bulk hollow cylinder. Our results agree with uni-axial stress experiments in literature, validating the bulk approximation. Here, we study the following configuration. The current is first ramp up to below the critical current and we calculate the screening currents and the forces that they cause using the MEMEP model. This electromagnetic model can now take insulated magnets into account. \R{As a case study 32 T REBCO superconductor magnet is simulated numerically}.  We have done complete mechanical analysis of the magnet by including the axial and shear mechanical quantities for each pancake unlike previous work where only radial and circumferential quantities are focused. Effect on mechanical quantities without screening current is also calculated and compared. It is shown that including screening current induced field strongly affect the mechanical quantities, specially the shear stress. The latter might be the critical quantity for certain magnet configurations. Additionally, in order to overcome high stresses, a stiff over banding of different material is considered and numerically modelled which significantly reduces the mechanical stresses. The FE based model developed is efficient to calculate the mechanical behaviour of any general superconductor magnet and its devices.
\end{abstract}

%
\vspace{2pc}
\ioptwocol
\section{Introduction}

 State of art high temperature superconducting REBCO tapes ($RE$Ba$_2$Cu$_3$O$_{7-x}$, where $RE$ is a rare earth like Y, Gd or Sm) are promising candidates for the development of future high field magnets \cite{liu2020world,hahn201945,weijers2013progress,cavallucci2019numerical,gavrilin2013observations}. These high field applications include magnets for materials research \cite{wikus2022commercial,yanagisawa2022review,fazilleau202038}, fusion reactors \cite{bruzzone2018high,mitchell2021superconductors,fietz2013prospects}, particle accelerators \cite{rossi2012superconducting,schmuser1991superconducting,cooley2005costs} and medical equipments \cite{alonso2012superconductivity,moser2017ultra,wang2023passive,vedrine2010iseult}. HTS windings has also been widely investigated to design electric air crafts \cite{sarlioglu2015more,luongo2009next,haran2017high,grilli2020superconducting}. The high field application require conductors with almost zero resistivity and extremely high current density \cite{salama1989high}. In such high magnetic field application there are high electro-mechanical forces and thus concern for mechanical failure \R{particularly degradation of REBCO coils.} In addition, screening current induced field \cite{shi2022numerical} as well as an-isotropic and in-homogeneous \cite{miyazaki2015delamination}  nature  of REBCO tape altogether connive to complicate the mechanics of deformation very significantly. Therefore it is important to develop a numerical tool to quickly and efficiently analyze the distribution of stresses and strain in high field REBCO coils.
  
  In recent years, a large number of numerical analyses are proposed to model the mechanics of REBCO superconductor coil insert for high field magnets. These works can be widely divided in two approaches. First approach for solving stress,strain and deformation is analytical as either in \cite{arp1977stresses,wilson1983superconducting,middleton1968mechanical} which is relatively straight forward where stresses are product of current density, coil radius and magnetic field or in \cite{gray1977electromechanical} where a continuum model is proposed neglecting the height of coil assuming the problem as plane stress. 
  
  The other and mostly used approach is implementation of commercial software numerical model such as \cite{niu2021numerical,wang2016experiment,ueda2022experiment,trillaud2022electromagneto,jing2020numerical,pi2021strain,xia2019stress,gao2020stress} and using transversely isotropic material properties for REBCO material. 
  
  Recently, based on aforementioned  numerical model Trillaud \textit {et al} \cite{trillaud2022electromagneto} discussed about effect of mechanical degradation on critical current density $J_c$. In another study Xia \textit{et al} \cite{xia2019stress}  shows the effect of shielding current on high filed REBCO Coils.  Ze Jing \cite{jing2020numerical} explored about the effect of phase field magnetization assuming the magnet as bulk. In another work, Ueda \textit{et al} \cite{ueda2022experiment} has shown the combined effect of cool down induced stresses and screening current on REBCO coil.
   A possibility of non circular pancake coils proposed by Wang \textit{et al} \cite{wang2014development}  indicates that this design can prevent degradation in REBCO wire due to deformation.
   
   In all the above study, subtle efforts have been made for mechanical modelling of REBCO coils on multi-physics software and couple it with electromagnetic forces in either sequential or iterative manner.

    Apart from the issue to solve this multiphysics problem numerically, for the safe design of high field magnet and devices, generation of screening current induced field (SCIF) \cite{pardo2016modeling}, is a challenge as it affects not only the mechanical behaviour \cite{wang2019dipole} but also homogeneity of the field \cite{maeda2013recent,pardo2016modeling}.

    To the best of our knowledge, there has not yet been a comprehensive study to solve mechanical equilibrium equations for an-isotropic REBCO superconductors magnet with SCIF and overbanding using finite element analysis. Recently, Yufan yan \textit{et al} \cite{yan2021screening} also discussed about impact of screening current and overbanding on REBCO magnets but the study limited to circumferential stresses and strains.

 In this study, we have developed a finite element scheme to study screening current induced stresses and strains in a high temperature axis-symmetric superconductor REBCO insert pancake coils. The strains are considered in linear elastic range for the current study. This is a reasonable assumption as for fail safe design of high field superconductor magnet firstly because the mechanical deformation should be in the range of reversible elastic range so that tapes are not degraded \cite{barth2013degradation} and secondly as the REBCO material is highly brittle \cite{hannachi2022mechanical} in nature. The finite element (FE) code is developed in MATLAB and coupled to already developed MEMEP model \cite{pardo2015EM,pardo2016modeling}. We rely on homogenization approach to deal with anisotropic nature of HTS tapes \cite{patel2015magnetic,xia2019stress}.
 
 The paper is organized as follows. In Section 2, the HTS insert geometrical and input parameters are described. In section 3, we have discussed the numerical model used to develop the computational framework to study the electro-mechanical analysis. 
 
 First, we present the electromagnetic model based on variational principles that calculates the screening currents and Lorentz forces. Later, we discuss the electromechanical model for axi-symmetric magnet geometry. A numerical fit shows the validation of bulk approximation with experimental uniaxial stress-strain. In that section, we introduce overbanding and the numerical scheme to handle its mechanics. Section 4 presents computation of mechanical properties with and without screening current and effect of overbanding for a 32T magnet.

\section{HTS magnet insert} \label{magnet parameters}
\begin{figure*}
     \centering
    \includegraphics[scale=0.18]{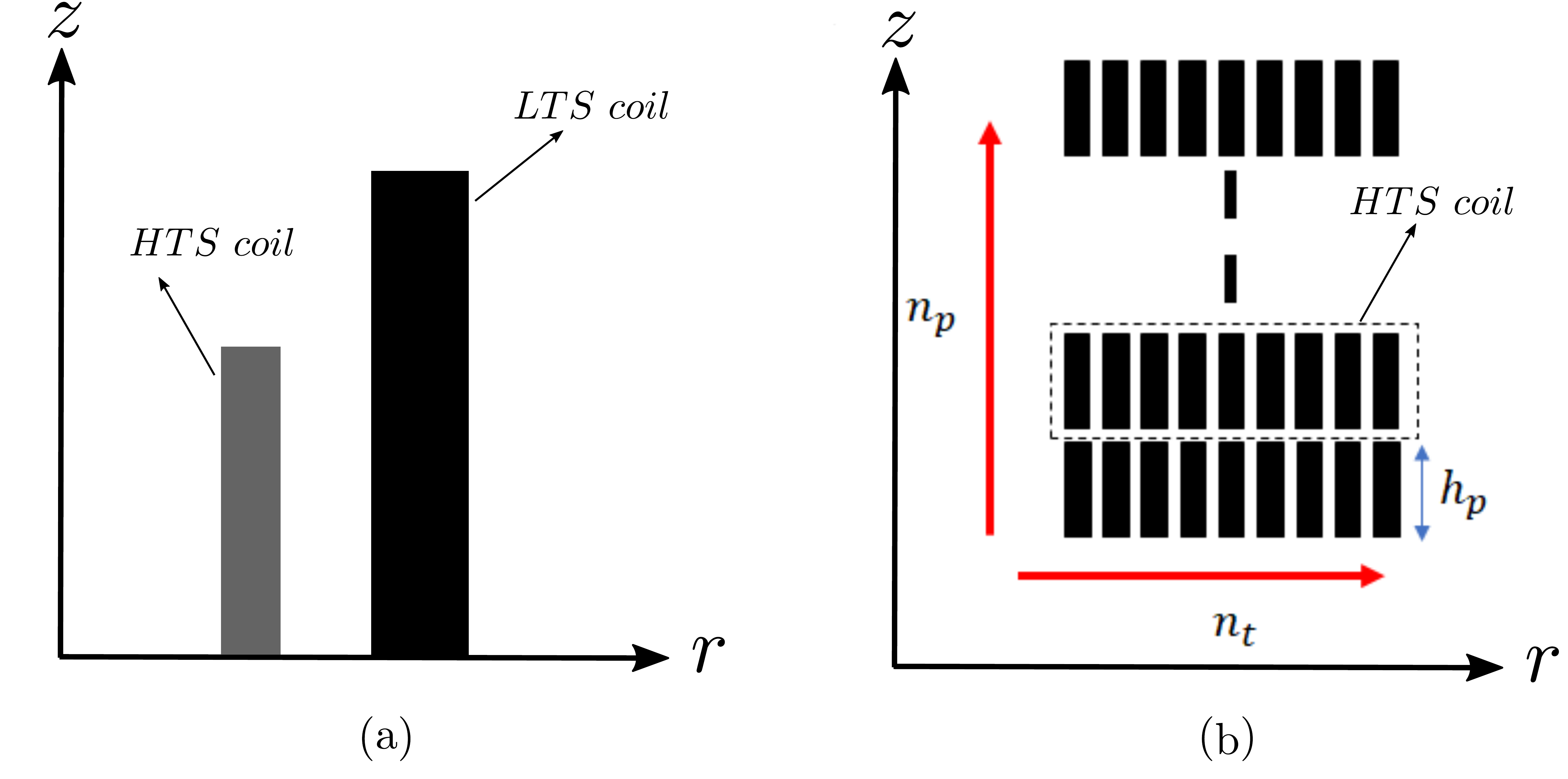}
     \caption{(a) Schematic of an Axis-symmetric REBCO Superconductor magnet in quarter $r$-$z$ plane and (b) HTS coil magnet with number of pancakes $n_p$ in z direction and number of turns $n_t$ in radial direction. The height of each pancake is $h_p$.}
      \label{HTS_magnet}
\end{figure*}

    An 8 double-pancake REBCO insert for a 32 T superconductor magnet is chosen as an example to implement mechanical properties. The REBCO insert is used in conjunction with LTS outer magnet which provide 19 T background field at the bore center. The magnet geometry and properties is given in figure \ref{HTS_magnet} and table \ref{magnet_prop}. 

\begin{table}
\begin{center}
\begin{tabular}{c|c} \hline
\multicolumn{2}{c}{32 T magnet coil geometry} \\ \hline
Parameter   &Value \\ \hline
$R_{in}$   & $25 \; \mathrm{mm}$ \\ 
$R_{out}$  & $56.5 \; \mathrm{mm}$ \\ 
Number of double pancake ($n_p$)  & $8 $ \\ 
Number of turn in each pancake ($n_t$)  & $250 $ \\ 
thickness of each tape layer & $105 {\mu m}$ \\ 
Height of pancake coil ($h_p$)& $6 \; \mathrm{mm}$ \\ \hline
\end{tabular}
\end{center}
\caption{Parameters used for REBCO insert magnet.}
\label{magnet_prop}
\end{table}

The REBCO insert design is for a metal-insulated coil, where a stainless-steel (SS) tape is co-wound with the superconductor. Taking into account, the high turn-to-turn surface resistance through the SS (around $10{^{-6}} - 10^{-7}$ $\Omega$m$^2$), the radial currents can be neglected \cite{pardoE2024SSTa}. \R{Thus, the tapes can be treated as electrically insulated}.

 The elastic and geometric properties are listed in table \ref{tape_prop}.

\begin{table}
\begin{center}
\begin{tabular}{c|c|c|c} \hline
    \multicolumn{4}{c}{Material properties of tape at 4.2 K} \cite{xia2019stress} \\\hline
Component   & $E$ & $\nu$ & t \\
            & (GPa) &  & ($\mu$$\mathrm{m}$) \\\hline
REBCO        & 180 & 0.3 & 1 \\ 
Hastealloy   & 210 & 0.3 & 80 \\ 
Copper       & 40 & 0.3 & 20 \\
Ag+Buffer    & 90 & 0.3 & 2 \\
SS tape & 156 & 0.3& 2 \\\hline
\end{tabular}
\end{center}
\caption{Material properties for REBCO tapes.}
\label{tape_prop}
\end{table}

\section{Numerical model}

\subsection{Electromagnetic model}

\begin{figure}
     \centering
    \includegraphics[scale=0.4]{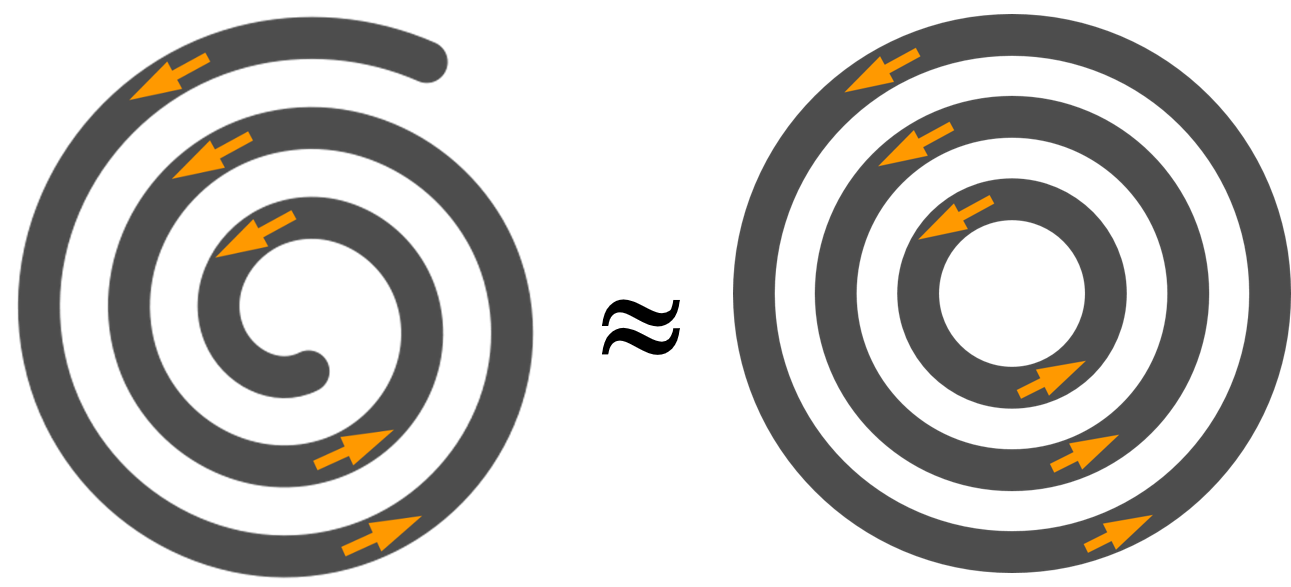}
     \caption{Spiral geometry of REBCO Superconductor coil winding and approximated as closely packed co-centric circular geometry. The arrow denotes the direction of flow of current for an insulated case.}
      \label{circular_approx}
\end{figure}

In this work, we compute the local current density taking screening currents into account by the Minimum Electro-Magnetic Entropy Production (MEMEP), as described in \cite{pardo2015EM,pardo2016modeling,pardoE2024SSTa}. For simplicity, we assume axi-symmetrical goemetry for the pancake coils (see, figure \ref{circular_approx}), which is justified because the tape and insulation thickness is much smaller than the coil inner radius. \R{Summarizing, MEMEP obtains the same kind of information as a finite-element method. However, MEMEP obtains the current density $\vJ$ at time $t$ by minimizing the following functional \cite{pardo20173d}
\begin{eqnarray}
F[\vJ] & = & \int_\Omega \dvol \bigg\{ \frac{1}{2} \Delta\vJ \cdot \frac{\vA[\Delta\vJ]}{\Delta t} + \Delta\vJ\frac{\Delta \vA_a}{\Delta t} + U(\vJ) \nonumber \\
&& + \Delta \phi\cdot \vJ  \bigg\}
\end{eqnarray}
provided that the current density at a previous time $t-\Delta t$, $\vJ(t-\Delta t)$ is known, where $\Delta t$ is a certain time step. At the equation above \R{$\Omega$ is the region where the conductors and superconductors exist}, $\vDJ\equiv \vJ(t)-\vJ(t-\Dt)$, $\vA[\vDJ]$ is the vector potential in the Coulomb's gauge that $\vDJ$ generates \cite{grilliF2014IES, pardoE2023book}, $\vA_a$ is the vector potential in Coulomb's gauge created by external sources (or applied vector potential), $\phi$ is the electrostatic potential, and $U(\vJ)$ is the loss factor, defined as
\begin{equation} \label{U}
U(\vJ)\equiv \int_0^\vJ \dif\vJ'\cdot\vE(\vJ'),
\end{equation} 
which is valid for non-linear $\vE(\vJ)$ relations of the material. The line integral of (\ref{U}) does not depend on the integration path for any physical $\vE(\vJ)$ relation, as shown in \cite{pardo20173d}. In this work, we assume the initial condition of $\vJ=0$ at $t=0$.

}

Regarding the electromagnetic properties, we assume that the superconducting layer presents a power-law relation between the electric field, $\vE$, and the current density $\vJ$ as
\begin{equation}
\vE(\vJ)=E_c\left ( \frac{|\vJ|}{J_c} \right )^n \frac{\vJ}{|\vJ|} , 
\end{equation}
where $E_c=10^{-4}$ V/m, $J_c$ is the critical current density, and $n$ is the power-law exponent. In this article, we assume that $n$ is constant with value 30 and $J_c$ depends on the magnetic field and its orientation, $J_c(B,\theta)$. As input, we use $J_c(B,\theta)$ according to the analytical fit of experimental data in \cite{fleiterJ2014CERN} for a Fujikura tape. We also consider the resistance of the metal stabilization, substrate and isolation by taking a homogenized anisotropic resistance into account, as in \cite{pardoE2024SSTa}. \R{In essence, this homogenization assumes the the superconducting layer and the metals are connected in parallel for the angular current density, $J_\varphi$, while they are connected in series for the radial current density, $J_r$. 

In this work, we assume $J_r=0$ for simplicity (non-insulated coil). The reason is that the screening currents after several minutes of charging the magnet in metal-insulated and non-insulated windings is the same as in insulated \cite{pardoE2024SSTa}. For axi-symmetric problems with insulated coils and current constraints, it is sufficient to minimize the following functional \cite{pardo2015EM}
\begin{eqnarray}
F[J_\varphi]& = & \int_{\Omega_s}\dif s 2\pi r \bigg\{ \frac{1}{2} \Delta J_\varphi \cdot \frac{A_\varphi[\Delta J_\varphi]}{\Dt} + \Delta J_\varphi \frac{\Delta A_{a,\varphi}}{\Dt} \nonumber \\
&& + U(J_\varphi)
\bigg\},
\end{eqnarray}
where $\Omega_s$ is the cross-section of the whole coil and $\dif s$ is the surface differential in the cross-section.

}

In this work, we consider that the radial turn-to-turn resistance is very high, and hence we assume that this resistance is infinite. In order to speed-up the computations, we group several turns in a single element in the radial direction (10 turns in this case).

In our model, we consider the realistic background magnetic field, $\vB_a$, created by the low-temperature superconducting (LTS) outsert, which we compute from the actual cross-section of the LTS winding and the Biot-Savart law. The generated magnetic field from the LTS at the bore center, as well as most of the HTS insert, is 19 T.

Once MEMEP finds both local current density and the magnetic field, the Lorenz force density is computed as
\begin{equation}
\mb{f}_L=\vJ\times\vB
\end{equation}
and in expanded general form
\begin{equation}
\mb{f}_L = {f}_r\mb{e}_r + f_{\phi}\mb{e}_{\phi} + f_z\mb{e}_z , 
\end{equation}
where $\mb{e}_r$, $\mb{e}_{\phi}$ and $\mb{e}_z$ are the unit vectors in radial, circumferential and axial directions and the components of the force density are ${f}_r = J_{\phi}B_z-J_{z}B_{\phi} $, ${f}_{\phi} = J_{z}B_r-J_{r}B_{z}$ and ${f}_z = J_{r}B_{\phi}-J_{\phi}B_{r}$. For isolated windings, $J_r$ vanishes and for tape an isolation thickness much lower than the inner radius, $J_z$ is negligible compared to $J_\phi$. Then, only $J_\phi$ is relevant, and hence
\begin{equation}
\mb{f}_L = J_{\phi}B_z\mb{e}_r -J_{\phi}B_{r}\mb{e}_z .
\label{F_density}
\end{equation}

As shown in \cite{pardoE2024SSTa}, the radial currents are negligible in metal-insulated magnets under regular operation. Thus, we can assume that the Lorentz forces in metal-insulated magnets is the same as in insulated.

\subsection{Electro-mechanical model} \label{Emmodel}
For the case study of magnet and developing the tool, the following assumption are made
\begin{enumerate}
    \item Thermal stresses are neglected.
    \item There is no sliding between the turns, they are perfectly bonded to each other.
    \item \R{Residual stresses due to winding tension are neglected. These stresses arises due to pre-compress of the REBCO layers which can be used as a tool to counter the outer radial stress during energization}.
\end{enumerate}
The assumption of coil turns of REBCO HTS into concentric coils simplified the problem domain to study the problem as axi-symmetry. In this study, each pancake is assumed as axi-symmetric hollow ring (see,\cite{pardo2016modeling}) and the corresponding stresses and strains are calculated. We deal the an-isotropic material properties of the tape with an analogous bulk model (see section \ref{homomodel}).  We start with the general deformation field given as,

\begin{figure*}
     \centering
    \includegraphics[scale=.25]{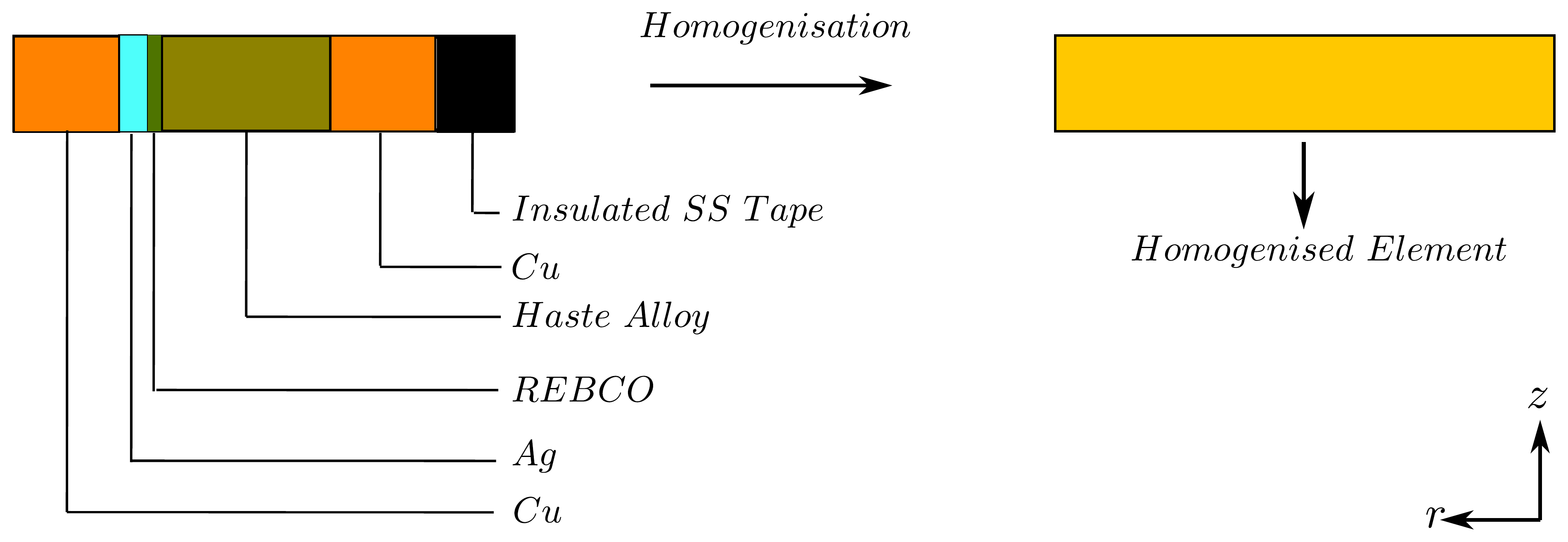}
     \caption{Schematic of tape layered structure and homogenised RVE of a single tape in r-z plane.}
      \label{homo}
\end{figure*}
\begin{equation}
    \mb{u} = u_r \mb{e}_r + u_{\phi} \mb{e}_{\phi} + u_z \mb{e}_z
\end{equation}

 and the corresponding strain tensor field is given by
 
\begin{equation}
    \mb{\epsilon} = \frac{1}{2} \left[ \mb{\nabla u + \nabla u}^T\right]
\end{equation}

due to axis-symmetry and the fact that circumferential displacement is independent of $\phi$ co-ordinate, the only strain components are
\begin{equation}
    \epsilon_r = \frac{\partial u_r}{\partial r}, \epsilon_{\phi} = \frac{u_r}{r}, \epsilon_z = \frac{\partial u_z}{\partial z}, \epsilon_{rz} = \frac{1}{2}\left(\frac{\partial u_r}{\partial z} + \frac{\partial u_z}{\partial r}\right)
    \label{strain_axis}
\end{equation}

Assuming the material within the range of linear elastic limit, the consistence stresses are related to strains by constitutive relationship  given as:
\begin{equation}
    \mb{\sigma} = D \mb{\epsilon}
    \label{stress-strain}
\end{equation}
where $D$ is the elasticity matrix, which presents anisotropic material properties due to the laminated structure of tape \cite{bower2009applied}. A bulk homogenization approach is used to calculate the components of elasticity matrix whose details are given in section \ref{homomodel}.
From the basic continuum mechanics the equilibrium equation for a body subjected to only body forces $\mb{f_b}$ is given by
\begin{equation}
   \mb{ \nabla \cdot \sigma} + \mb{f_b} =0
\end{equation}
Due to axis-symmetry geometry,  the only stress components are $\sigma_r$, $\sigma_{\phi}$, $\sigma_z$ and $\sigma_{rz}$ and the governing equations are given by

\begin{equation}
    \frac{\partial {\sigma_r}}{\partial r} +  \frac{\partial {\sigma_{rz}}}{\partial z} + \frac{\sigma_r - \sigma_{\phi}}{r}+f_r =0, \nonumber 
\end{equation}
\begin{equation}
    \frac{\partial {\sigma_{rz}}}{\partial r} +  \frac{\partial {\sigma_{z}}}{\partial z} + \frac{\sigma_{rz}}{r} + f_z =0
    \label{governing_mech}
\end{equation}
 where $f_r$ and $f_z$ are body forces due to electro-mechanical Lorentz force in radial and axial directions respectively.

\begin{figure*}
     \centering
    \includegraphics[scale=0.55]{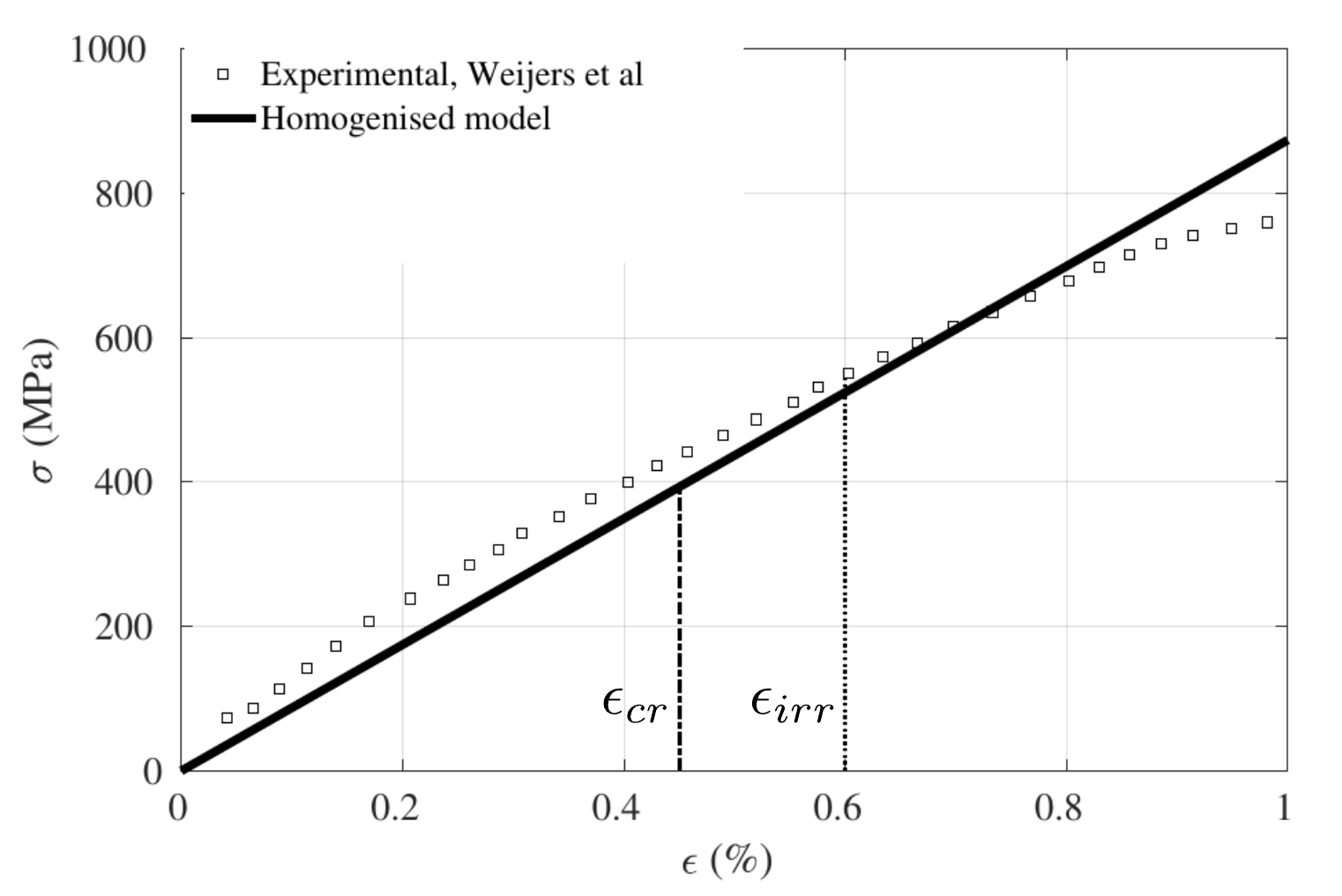}
     \caption{Homogenised model compared with experimental stress strain of REBCO Superconductor \cite{weijers2016progress}. The dash-dot vertical lines are correspond to $\epsilon_{cr}=0.45 \%$ below which critical current density $J_c$ of the tape is almost constant and $\epsilon_{irr}=0.6 \%$ \cite{barth2015electro} beyond which there is plastic deformation and critical current density $J_c$ will permanently degraded.}
      \label{expvsmodel}
\end{figure*}

\subsubsection{Homogenized Element}: \label{homomodel}   
 In this section, the anisotropy of the tape structure is dealt in a similar way as explained in \cite{xia2019stress}. This approach is known as bulk approximation and widely used  assuming the laminated pancake as a continuum disk. The equivalent homogeneous representative volume element (RVE)  have isotropic property in the $\phi-z$ plane (see, figure \ref{homo}) and different in radial direction. Further,  the transversely isotropic elasticity matrix ${D}$ which relate axis symmetric stresses with corresponding strains is given by

\begin{equation}
    \begin{bmatrix} \sigma_r \\ \sigma_{\phi} \\ \sigma_z \\ \sigma_{rz} \end{bmatrix} =  \begin{bmatrix} D_{11}&D_{12}&D_{12}&0 \\ D_{12}&D_{22}&D_{23}&0 \\ D_{12}&D_{23}&D_{22}&0 \\ 0&0&0&D_{44}\end{bmatrix} 
    \begin{bmatrix} \epsilon_r \\ \epsilon_{\phi} \\ \epsilon_z \\ \epsilon_{rz} \end{bmatrix}
\end{equation}

 The components of the $D$ matrix are a function of the Elastic modulus $\bar{E}$, Poisson's ratio $\bar{\nu}$ and thickness of each layer of laminated tape $t_m$ and calculated following \cite{bower2009applied}. The homogenised material properties are given as following
  \begin{equation}
      D_{11}=\bar{E}_{r} = \frac{1}{\sum_{m=1}^{n} t_m/E_m}  \nonumber
      \end{equation}
\begin{equation}
    D_{22}=\bar{E}_{\phi} = \bar{E_{z}} ={\sum_{m=1}^{n} t_m E_m} \nonumber
\end{equation}
      
      \begin{equation}
      \bar{\nu}_{\phi z} = \frac{{\sum_{m=1}^{n} t_m \nu_m E_m}}{{\sum_{m=1}^{n} t_m E_m}} , \bar{\nu}_{rz} = \frac{\bar{E}_r}{\bar{E}_{z}}  \sum_{m=1}^{n} t_mE_m \nonumber
  \end{equation}
  \begin{equation}
      D_{12}=D_{44}=\bar{\nu}_{rz} \bar{E}_{r}, D_{23}=\bar{\nu}_{\phi z}\bar{E}_{\phi}
  \end{equation}
where $n$ is number of laminated layer in REBCO tape, $E_m$ and $t_m$ are modulus of elasticity and thickness of the tape layer respectively.

We then calculated the $\bar{E}$ matrix from $E_m$ and $\nu_m$ experimentally obtained in \cite{weijers2016progress}. Clearly the model is able to capture the uni-axial stress-strain response as shown in figure \ref{expvsmodel}. At this stage it is more appropriate to say that with in elastic range as shown in  figure \ref{expvsmodel} the model response is quite good.

\begin{figure}
     \centering
    \includegraphics[scale=0.5]{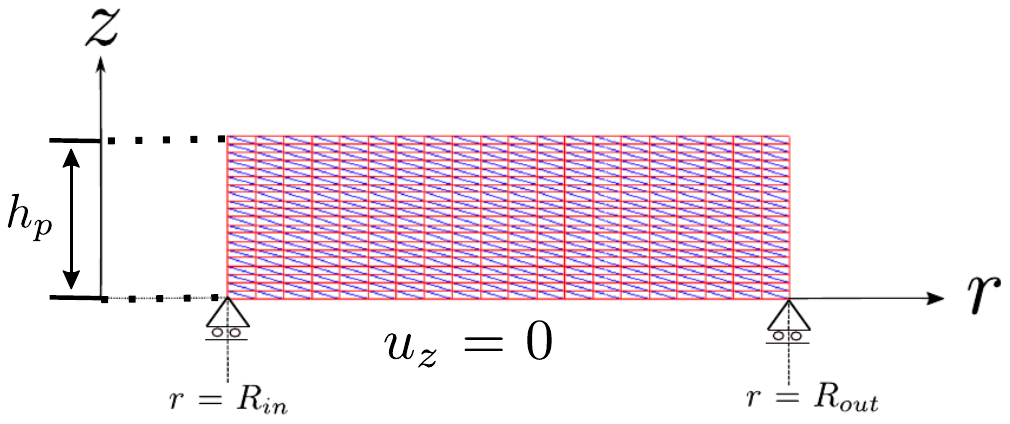}
     \caption{A typical mesh of a pancake  with triangular element. Each pancake in HTS magnet is subjected to roller boundary condition i.e. axial displacement $u_z$ = 0 on bottom of each pancake.}
      \label{mesh}
\end{figure} 

\subsubsection{Finite Element Implementation}: \label{FE Form}
  In this section we closely follow finite element formulation as in \cite{hutton2004fundamentals} for axis-symmetry space to obtain mechanical stress and strain for REBCO tapes subjected to electromechanical forces. The charging process of the magnet is assumed very slow so that inertia effects can be neglected. 
 In general finite element approach, the global stiffness $K_G$, nodal displacements and global nodal forces $F_G$ are related as 
 \begin{equation}
     [K_G] [U] =[F_G]
     \label{eqb_G}
 \end{equation}
 in an element they follow
  \begin{equation}
     [K_e] [U_e] =[F_e]
     \label{eqb_element}
 \end{equation}
 where $K_e$, $U_e$ and $F_e$ are element stiffness, nodal displacement and nodal forces in the element, respectively.

 Next we briefly discuss on finding the element stiffness and subsequently global displacement for the problem which is well defined (see \cite{hutton2004fundamentals} for detail).
 First we define radial and axial displacement field  ${u_r}$ and ${u_z}$ which are given by in discretized form as
\begin{equation}
    {u_r} = \sum_{i=1}^{n_e} N_i U_{r_i} \nonumber
\end{equation}

\begin{equation}
    {u_z} = \sum_{i=1}^{n_e} N_i U_{z_i}
\end{equation}

where $n_e$ is the number of node where $u_r$ and $u_z$ are calculated in an element and $N_i$ are shape functions. Though shape functions depend on the choice of element shape, general procedure will remain as same. 

\R{For the present study we have descretized the pancake domain in triangular elements as shown in figure \ref{mesh}. The $U_{r_i}$ and $U_{z_i} $ are the radial and axial displacement of $i_{th}$ node respectively.
 Assuming triangular elements with node indexes $i, j, m$, the axis-symmetric strains in each element} are given by

\begin{equation}
    \begin{bmatrix} \epsilon_r \\ \epsilon_{\phi} \\ \epsilon_z \\ \epsilon_{rz} \end{bmatrix} = \begin{bmatrix} \frac{\partial N_i}{\partial r}&0&\frac{\partial N_j}{\partial r}&0&\frac{\partial N_m}{\partial r}&0\\ \frac{N_i}{r}&0 &\frac{N_j}{r}&0&\frac{N_m}{r}&0\\  0&\frac{\partial N_i}{\partial z}&0&\frac{\partial N_j}{\partial z}&0&\frac{\partial N_m}{\partial z}\\ \frac{\partial N_i}{\partial z}&\frac{\partial N_i}{\partial r}&\frac{\partial N_j}{\partial z}&\frac{\partial N_j}{\partial r}&\frac{\partial N_m}{\partial z}&\frac{\partial N_m}{\partial r}\end{bmatrix} 
    \begin{bmatrix} U_{r_i} \\ U_{z_i} \\ U_{r_j} \\ U_{z_j} \\ U_{r_m} \\ U_{z_m} \end{bmatrix},
\end{equation}
 which can be written as
 \begin{equation}
     [\epsilon_e] = [B] [U_e]
 \end{equation}
 where the components of B are differential of shape function of the element and refer as the geometric stiffness as it depend on geometric parameters and it relates the strain with the displacement.

 Using equation (\ref{stress-strain}), elemental stresses are calculated as 
 \begin{equation}
     [\sigma_e] = [D] [B] [U_e]
 \end{equation}

 The strain energy due to deformation in an element is given by,
                \begin{equation}
                    W_{int} = \frac{1}{2}\int_{v_e} \mb{\sigma} \cdot \mb{\epsilon} dv 
                    \end{equation} 
                 Hence,    
                   \begin{equation} 
                   W_{int} = \frac{1}{2} [U_e]^T \left[ \int_{v_e} [B]^T  [D] [B]  dv \right] [U_e]
                   \end{equation}
In the finite element set up for solid materials the elemental stiffness $K_e$ is given by
 
 \begin{equation}
     K_e = \int_{v_e} [B]^T [D] [B] dv
 \end{equation}
 
  which can be modified for the axis symmetric case as
  
  \begin{equation}
     K_e = 2\pi \int_{dA_e} [B]^T [D] [B] rdrdz
 \end{equation}
  and further can be approximated as

  \begin{equation}
      K_e = 2\pi \bar{r}_e A_e [\bar{B}]^T [D] [\bar{B}]
  \end{equation}

  where $\bar{r}_e$ is radial coordinate of centroid of element and $\bar{B}$ are calculated at centroid of element. $A_e$ is the cross section area of element.
   The global stiffness $K_G$ is now calculated using component of elemental stiffness $K_e$.  

Now to calculate forces which for the present situation is only body forces and obtained as explained in section \ref{Emmodel}. Following finite element setup as explained in \cite{hutton2004fundamentals}

For small element size, the force vector at $i_{th}$ node is
   \begin{equation}
       F_i = \frac{2}{3}\pi \bar{r}_e A_e [f_L]\nonumber \\
\end{equation}
and in elemental
       \begin{equation}
       \begin{bmatrix}{F_e} \end{bmatrix} = {\begin{bmatrix} F_{i}\\F_{j}\\F_{k} \end{bmatrix}}
   \end{equation}

  where $f_L$ is lorentz forces density whose components are obtained from equation (\ref{F_density}). Each node has two degree of freedom in radial and axial direction respectively. The corresponding global force $F_G$ then can be calculated using elemental forces $F_e$.
  
At this stage, deformation or forces at the nodes can be calculated using global displacement and force relation (see, equation \ref{eqb_G}). Also,  since we are using first order triangular element so number of integration point is only the centriod of the triangle and strains as well as stresses are calculated at this integration point.

  \subsubsection{Overbanding Stiffness}:
  
  \begin{figure}
     \centering
    \includegraphics[scale=0.65]{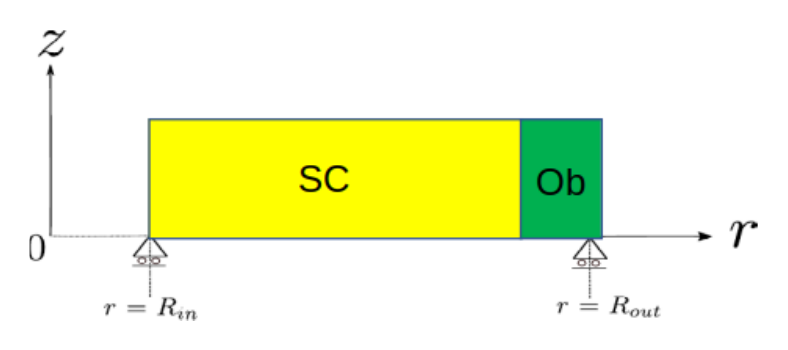}
     \caption{A schematic of pancake with over banding layers subjected to roller boundary condition in $r-z$ plane}
     
      \label{OB_schematic}
\end{figure}
   It is important to discuss the inclusion of stiffness property of the overbanding as shown in figure \ref{OB_schematic} for the problem at this stage. For the numerical analysis, the following assumption are made
\begin{enumerate}
    \item The overbanding turns are considered as continuum bulk.
     \item The adjacent overbanding turn are perfectly bonded to each other.
     \item The junction of superconductor tape and overbanding material share commom boundary and are perfectly bonded to each other. 
\end{enumerate}
  
 These assumptions do not enable slip between the two different materials: superconducitng coil and overbanding.
 
 The pancake is discretized in triangular elements as shown in figure \ref{mesh}. Now if total number of node in the superconducting material are $n_{t_{sc}}$ and total number of nodes in whole domain are $n_t$.

If value all the three nodes of the element is less than  $n_{t_{sc}}$, the elemental stiffness for the triangular element is given by, 
    
\begin{figure}
     \centering
    \includegraphics[scale=0.55]{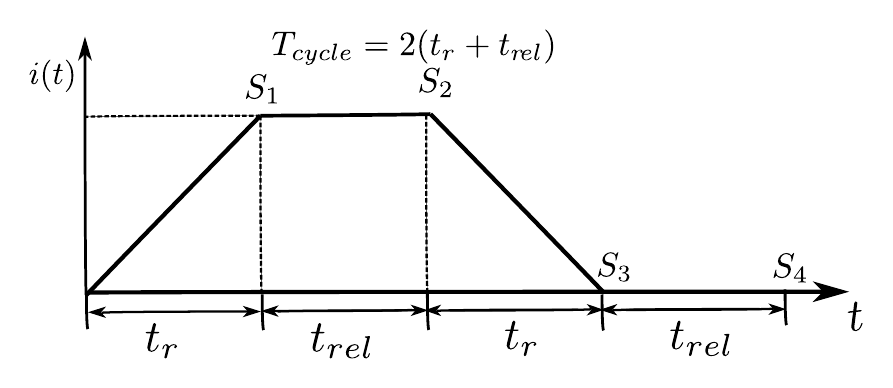}
     \caption{A Cycle of input current in a magnet used for charging and discharging the magnet.First the current is ramp up to $I_{max}$ corresponding to $S_1$ and till point $S_2$ this $I_{max}$ is kept constant. The current is then ramped down to zero during $S_2$ and $S_3$ and in this zero state for interval $S_3$ and $S_4$.} 
      \label{cycle_I}
\end{figure}

\begin{equation}
      K_e = 2\pi \bar{r}_e A_e [\bar{B}^T] [D_{sc}] [\bar{B}], \nonumber 
\end{equation}

    otherwise

    \begin{equation}
      K_e = 2\pi \bar{r}_e A_e [\bar{B}^T] [D_{ob}] [\bar{B}]
  \end{equation}

  where $D_{sc}$ and $D_{ob}$ are material properties for REBCO superconductor and overbanding material respectively.

\section{Results}

 The numerical model developed in section \ref{FE Form} is implemented to investigate the mechanical properties of the 32 $\mathrm{T}$ REBCO superconductor HTS magnet as explained in section \ref{Emmodel}. 
 
 Firstly, the pancake geometry and mesh as shown in figure \ref{mesh} is drawn in MATLAB using partial differential equation (PDE) tool module. The mesh properties nodes, triangles and edges are then imported to in-house developed tool programmed in MATLAB to further calculate the stiffness and mechanical properties. As shown in figure \ref{mesh}, the bottom of each pancake is subjected to roller boundary condition, and hence $z=z_{\rm bottom}$, where $z_{\rm bottom}$ is the axial coordinate of the bottom of the pancake.

 We consider a charge and discharge cycle of the magnet as shown in figure \ref{cycle_I}. There are four salient features of cycle. First during the charging, the current is ramped to $I_{max}$ for $t=t_r$ (point $S_1$) and the current $I(t)$ at any particular time of the cycle is given as

\begin{strip}
\begin{equation}
I(t) = \left \{ \begin{array}{cc}
I_{max} \dfrac{t}{t_r}  & \mbox{for $0 < t \leq t_r$} \\
I_{max}   & \mbox{for \; $t_r < t \leq t_r+ t_{rel}$} \\
-I_{max} \left ( \dfrac{t-t_{rel}}{t_r} - 2 \right ) 
&\mbox{for $t_r+ t_{rel} < t \leq 2t_r + t_{rel}$} \\
0   &\mbox{for $2 t_r + t_{rel} < t < 2(t_r + t_{rel})$},
\end{array}
\right .
\label{eq:tr}
\end{equation}
\end{strip}

where, $t_r$ is the ramp up as well ramp down charging time. The ramp down discharging time is also same. The relaxation time after charging and discharging is $t_{rel}$ . Then, the total time for one cycle is $T=2(t_r+t_{rel})$. For the current analysis $t_r=t_{rel}=333$ $\mathrm{s}$ and $I_{max}=333$ $\mathrm{A}$. Hence, the rate of ramp is 1 $\mathrm{A/s}$.

\begin{figure}
     \centering
    \includegraphics[scale=0.19]{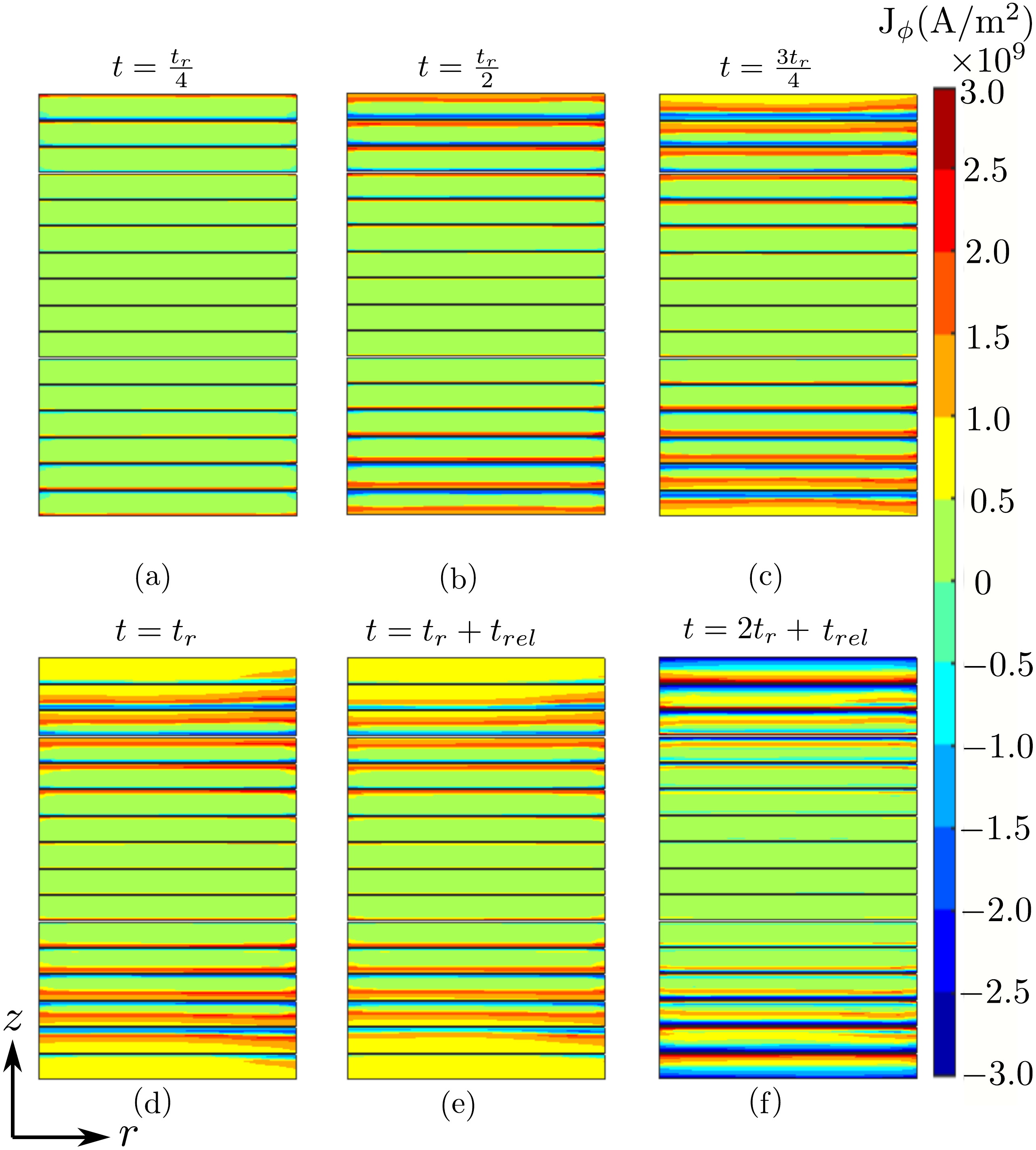}
     \caption{Variation of circumferential current density $J_{\phi}$ in a cycle for the whole magnet with screening current induced field. (a) to (c) is charging, (d) fully charged state, (e) discharged state and (f) discharged state with relaxation time $t_{rel}$ .}
      \label{Current density}
\end{figure}

\subsection{Electromagnetic analysis}  \label{results_EM}
In this section, we have done electromagnetic analysis using the inputs defined in section \ref{magnet parameters} and using the coil geometry given in table \ref{magnet_prop}. First, we discuss the initial ramp (figure \ref{Current density}(a-d)). The circumferential current density $J_{\phi}$ is shown for each pancake at every quarter of ramp time $t_r$. Initially, at low current the current density is uniform and low in all pancakes, except a narrow layer close to the top and bottom of each pancake but as we increase the $I$, the current density is high in most of the pancakes cross-section, specially in the upper and lower pancakes.  This is due to time-varying magnetic field in the radial direction. Initially, for $t\le t_r/2$, the current density is roughly anti-symmetric with respect to the center of each pancake cross-section.

The magnet is subjected to constant current and then it is decreased in discharging process. The current density for these cases are in figures \ref{Current density}(e) and \ref{Current density}(f), respectively. In the latter, the current density is purely due to screening currents.

\begin{figure}
     \centering
    \includegraphics[scale=0.19]{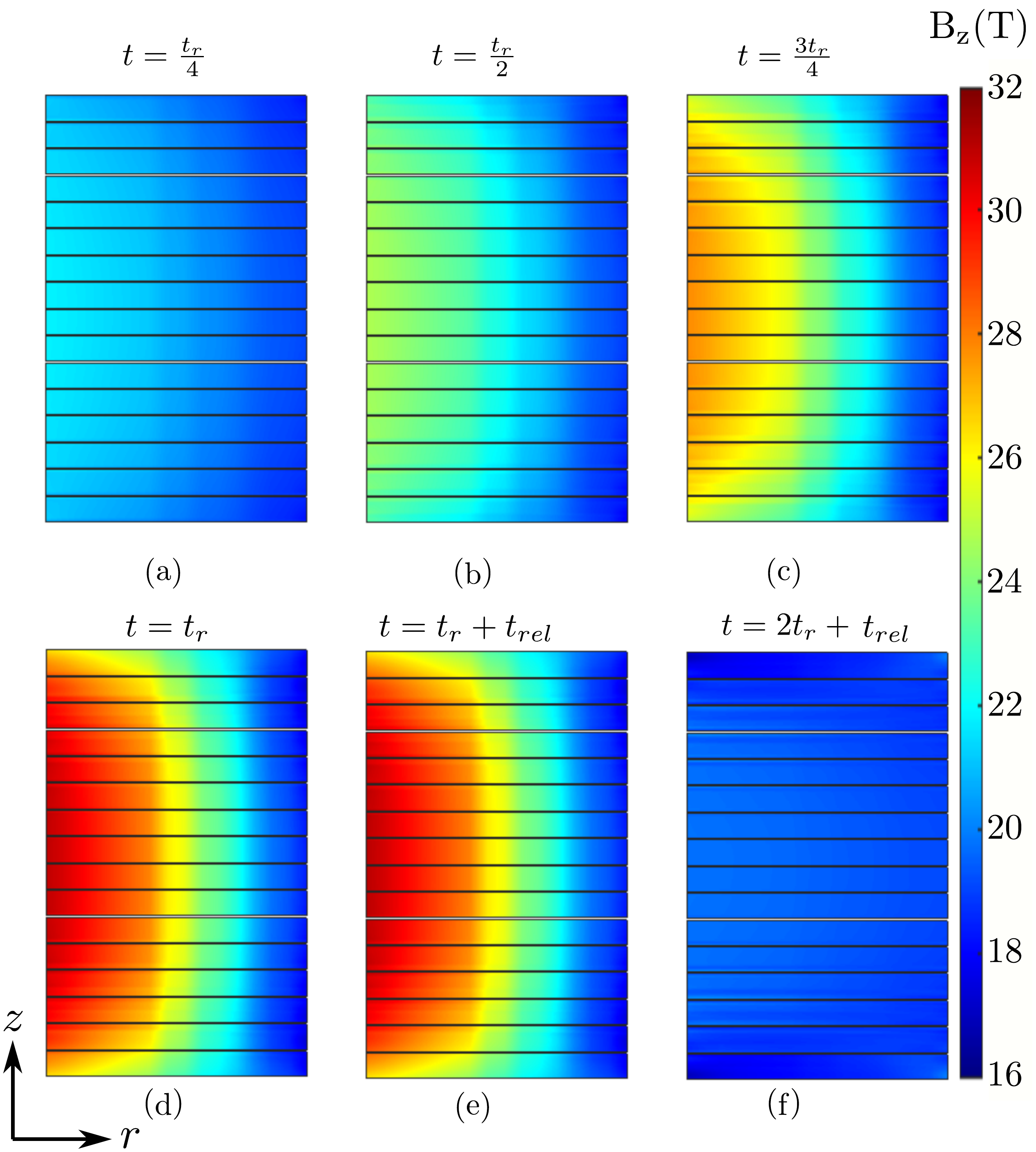}
    \caption{Variation of axial magnetic field $B_{z}$ in a cycle for the whole magnet with screening current induced field. (a) to (c) Charging, (d) fully charged state, (e) discharged state and (f) discharged state with relaxation time $t_{rel}$ .}
      \label{axial_Bz}
\end{figure}

\begin{figure}
     \centering
    \includegraphics[scale=0.19]{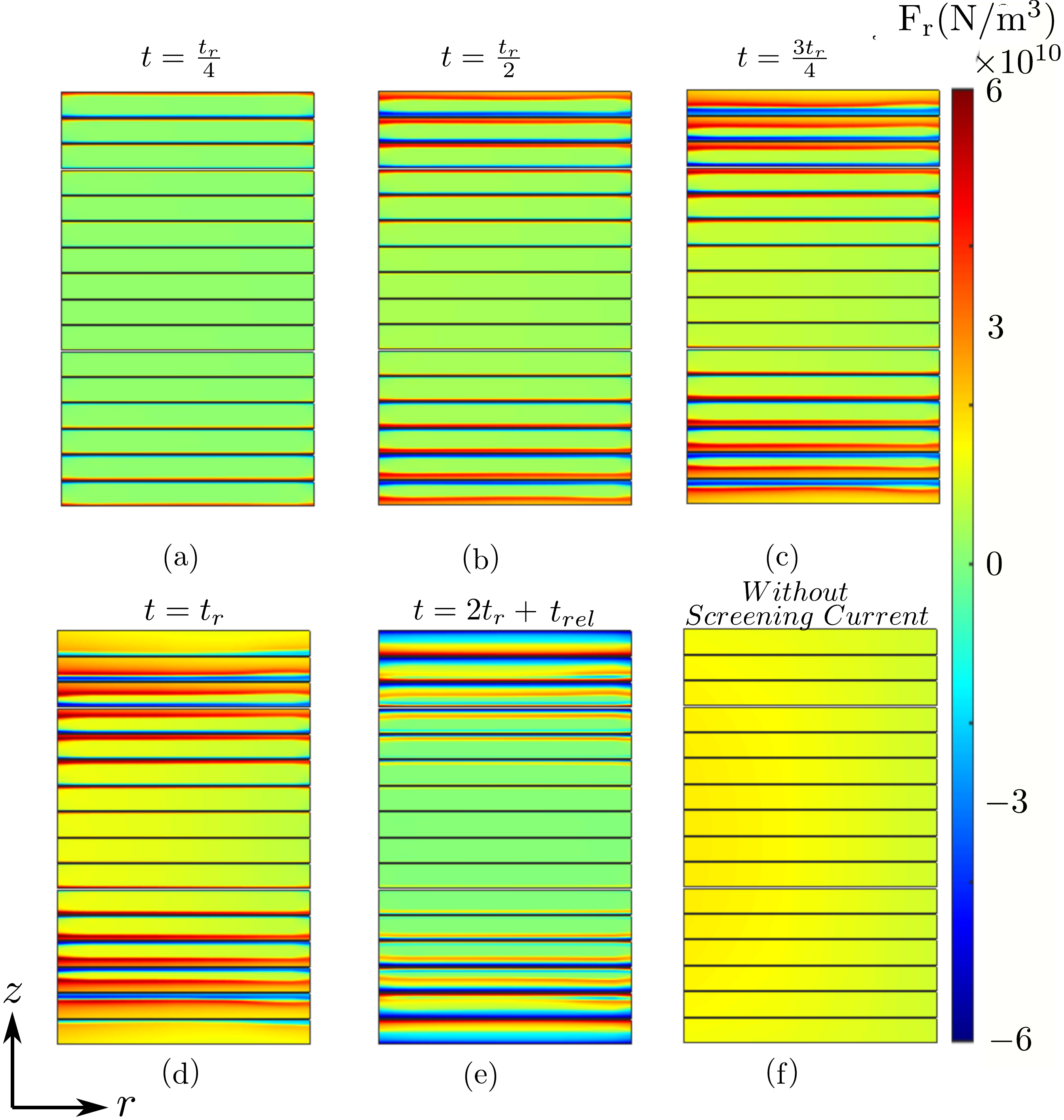}
    \caption{Variation of radial force in a cycle for the whole magnet with screening current induced field. (a) to (c) is charging, (d) fully charged state, (e) discharged state and (f) without Screening current \R{at time $t=t_r$}}
      \label{radial_force}
\end{figure}

The axial magnetic field is shown for the same process in figure \ref{axial_Bz}. The field monotonically decreases with the radius. Also, the maximum magnetic field, of around 32 T, is at the inner radius of the central pancakes.

Figure \ref{radial_force}(a-e) shows the radial Lorentz force density. Screening current effect can be easily seen as pancakes farther from center experience positive and negative radial forces within the same pancake. The magnitude of maximum force density is $6 \times 10^{10}$ N/m$^3$.  We have also analyzed the force in the magnet at the end of the ramp when there is no screening current (see, figure \ref{radial_force}(f)). Clearly, forces are all positive, as well as more uniform and lower than for the case without screening current, reaching a maximum value 3 times lower than when screening currents are taken into account. The lower value of the force is due to lower local current density.

The axial forces in the magnet are much lower ($\approx$ $\frac{1}{10}$ times) than radial forces, and hence they effectively do not contribute to magnet mechanics. This is due to the fact that the axial magnetic field ${B}_z$ is much greater than the radial magnetic field ${B}_r$.

\begin{figure}
     \centering
    \includegraphics[scale=0.19]{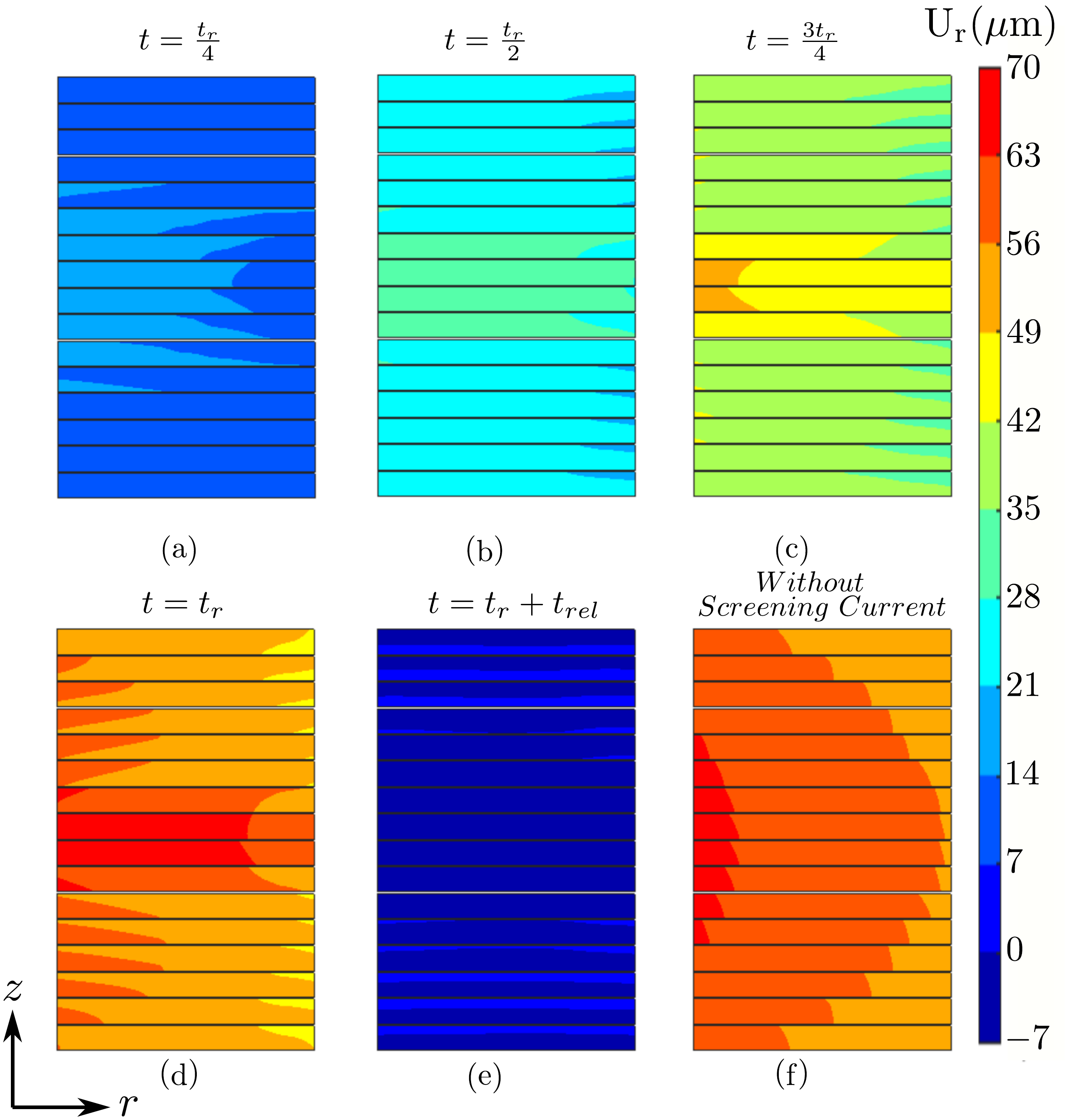}
    \caption{Variation of radial displacement $u_r$ in a cycle for the whole magnet with screening current induced field. (a) to (c) Charging, (d) fully charged state, (e) discharged state and (f) without Screening current \R{at time $t=t_r$}.}
      \label{displacment}
\end{figure}

\begin{figure}
     \centering
    \includegraphics[scale=0.19]{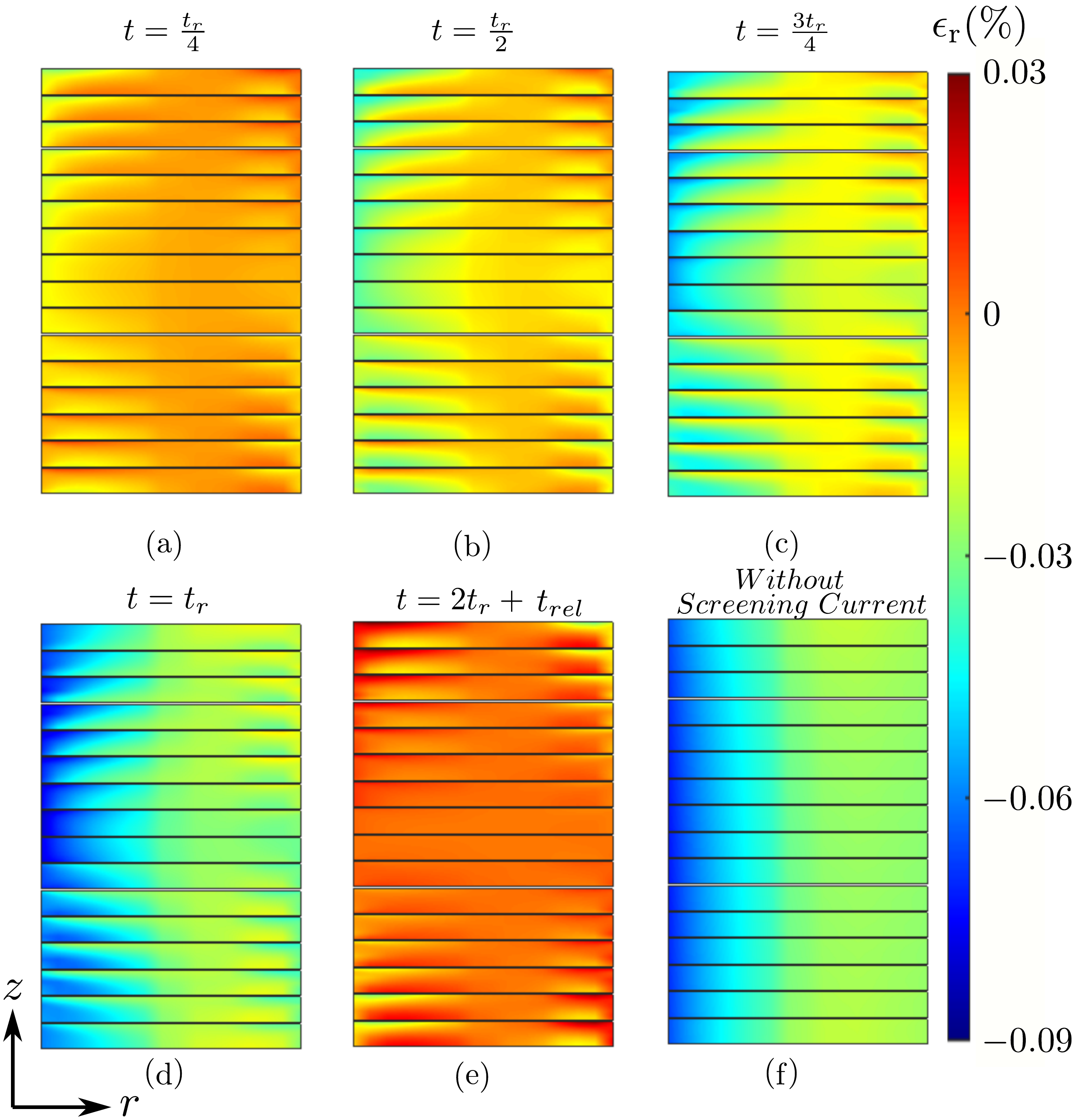}
    \caption{Variation of radial strain $\epsilon_{r}$ in a cycle for the whole magnet with screening current induced field. (a) to (c) Charging, (d) fully charged state, (e) discharged state and (f) without Screening current \R{at time $t=t_r$}.}
      \label{radial_strain}
\end{figure}

\begin{figure}
     \centering
    \includegraphics[scale=0.185]{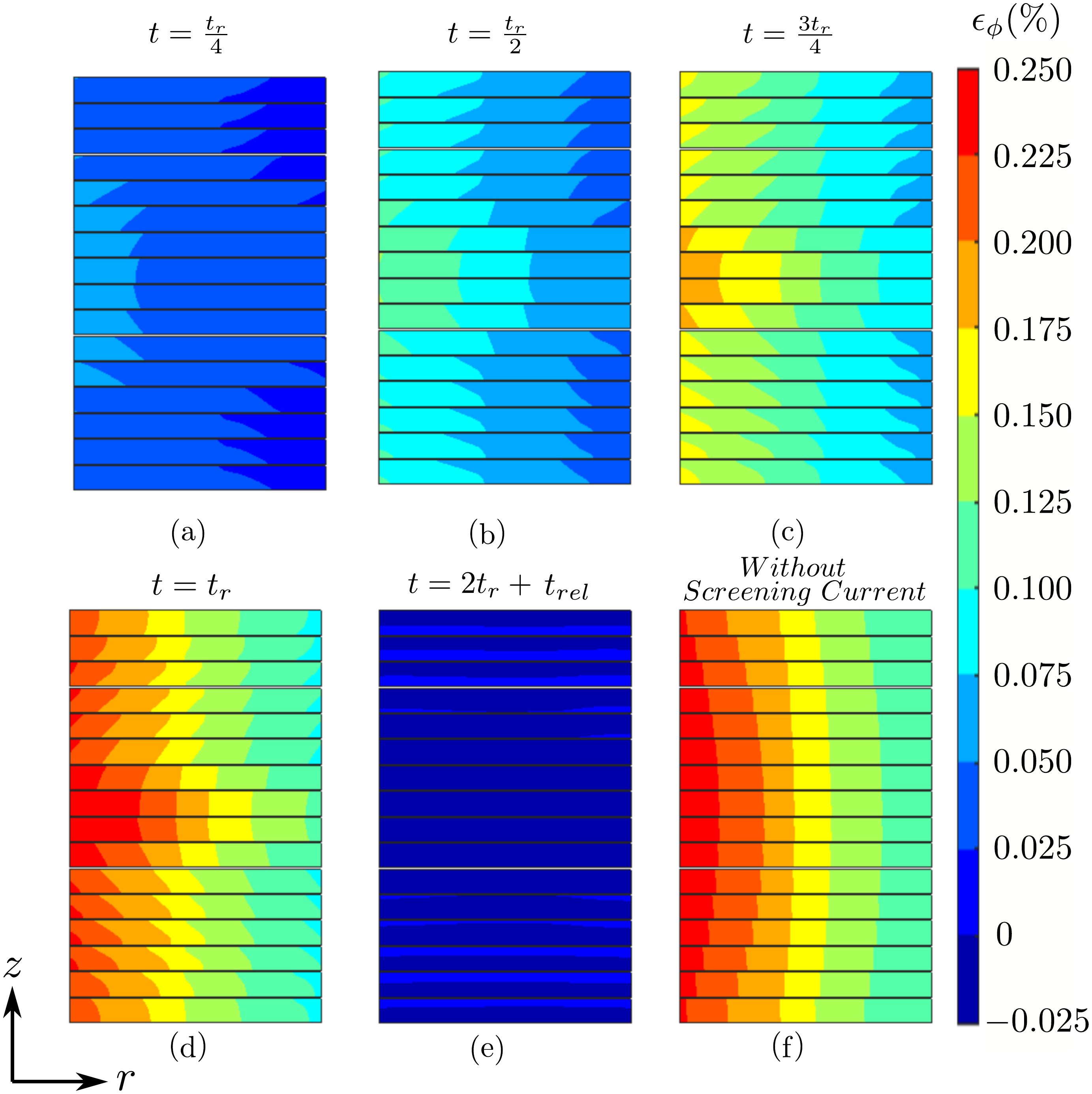}
    \caption{Variation of circumferential strain $\epsilon_{\phi}$ in a cycle for the whole magnet with screening current induced field. (a) to (c) Charging, (d) fully charged state, (e) discharged state and (f) without Screening current \R{at time $t=t_r$}.}
      \label{phi_strain}
\end{figure}

\begin{figure}
     \centering
    \includegraphics[scale=0.19]{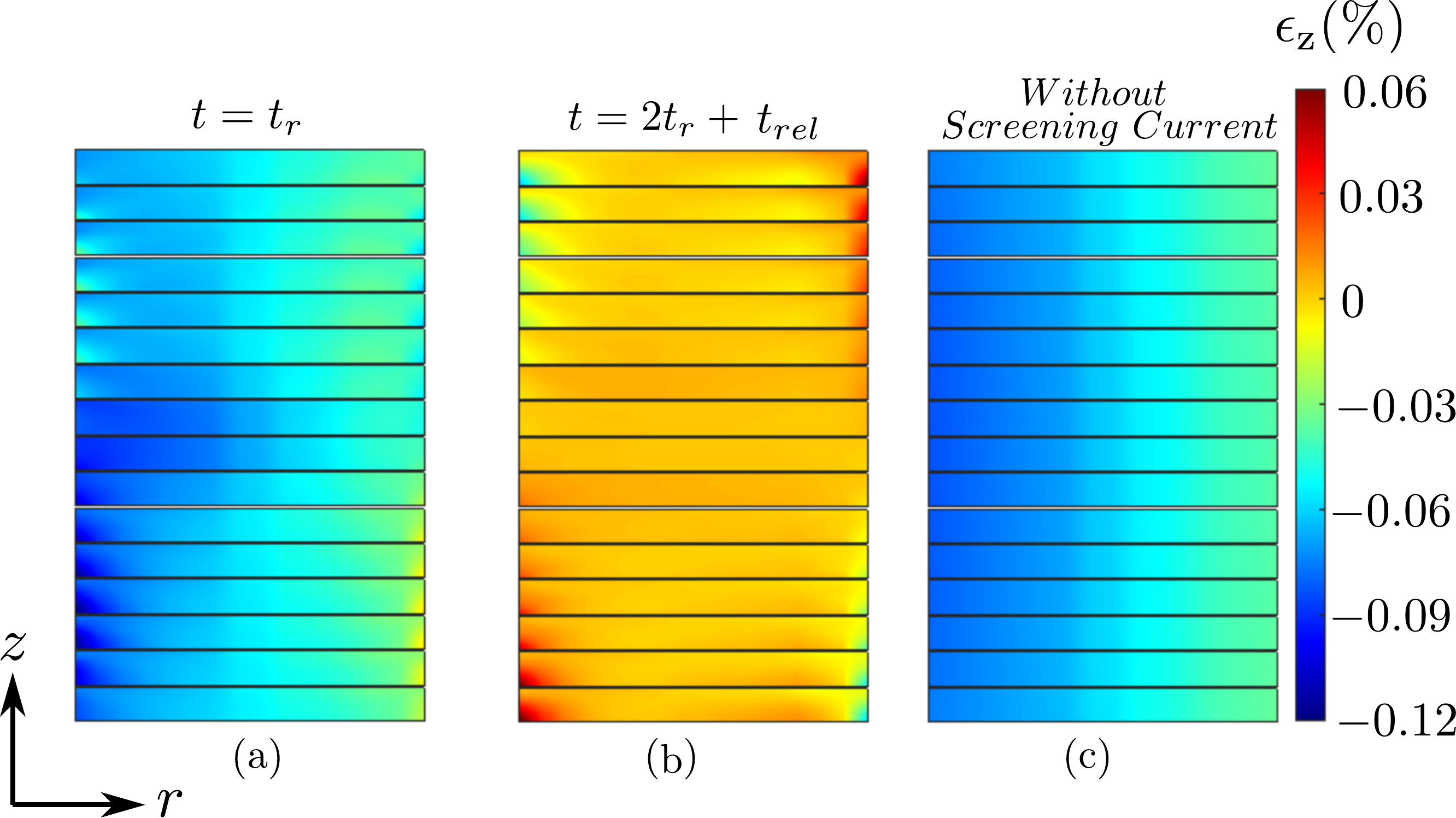}
     \caption{Variation of axial strain $\epsilon_{z}$ * for the whole magnet with screening current induced field. (a) fully charged state, (b) discharged state and (c) without Screening current \R{at time $t=t_r$}.}
      \label{axial_strain}
\end{figure}

\subsection{Mechanical response with and without Screening current}
We now import forces to the finite element model to further calculate the mechanical quantities in sequential manner.

\subsubsection{Displacement and strain fields in magnet:}First, we calculate the radial displacement (see, figure \ref{displacment}). The maximum displacement occurs at the maximum current and at center of the pancakes in magnet because the radial forces there are all positive (see, figure \ref{radial_force}(d)), and hence these forces are pushing the superconductor radially outwards. The maximum magnitude of displacement is only $70$ $\mu$m. Till this point, the displacement is radially outward only. Interestingly, when the current is ramped down in discharged state (see,  figure\ref{displacment}(e)), the deformation field is highly non homogeneous due to forces that are also non-uniform because of the screening currents. If we observe the top of the pancake at the outermost diameter, the deformation is negative and positive, that is on the edges the deformation is not uniform.

The displacement field when there is no screening current is maximum at the midplane (figure \ref{displacment}(f)). The comparison of the two cases clearly shows that the maximum displacement is highly non-uniform in the magnet subjected to \R{screening current induced field(SCIF)}.

\begin{figure}
     \centering
    \includegraphics[scale=0.19]{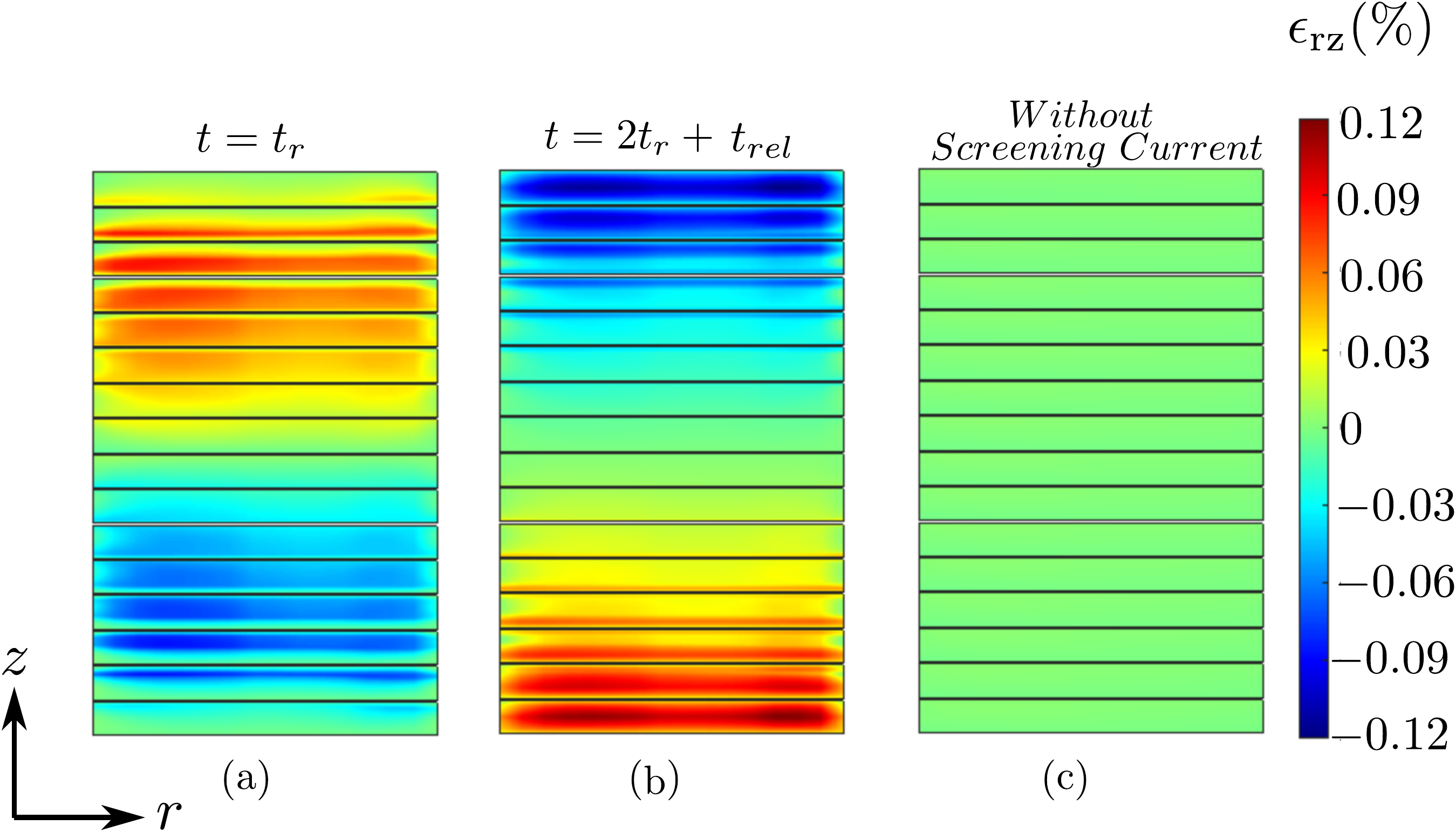}
    \caption{Variation of shear strain $\epsilon_{rz}$ in a cycle for the whole magnet with screening current induced field. (a) fully charged state, (b) discharged state and (c) without Screening current \R{at time $t=t_r$}.}
      \label{shear_strain}
\end{figure}

Figures \ref{radial_strain}-\ref{axial_strain} show the radial, circumferential and axial strain in the magnet. 

\R{The radial strain field is compressive} while charging, whether screening current is considered or not. The maximum magnitude of radial strain is $0.09$ $\%$ when the magnet is fully charged. The maximum strain appears in the central pancakes (see figure \ref{radial_strain}(d)). For the pancakes farther from the center, one may clearly see that the edges are subjected to non uniform radial strain due to SCIF. At fully discharged state $\epsilon_r$ is both tensile and compressive mostly in the top and bottom pancakes (see figure \ref{radial_strain}(e)). The variation is from tensile $+0.03$ $\%$ to compressive $-0.03$ $\%$ within the pancake. For no screening currents at the end of the ramp, the radial currents are more homogeneous (figure \ref{radial_strain}(f)).

\begin{figure}
     \centering
    \includegraphics[scale=0.19]{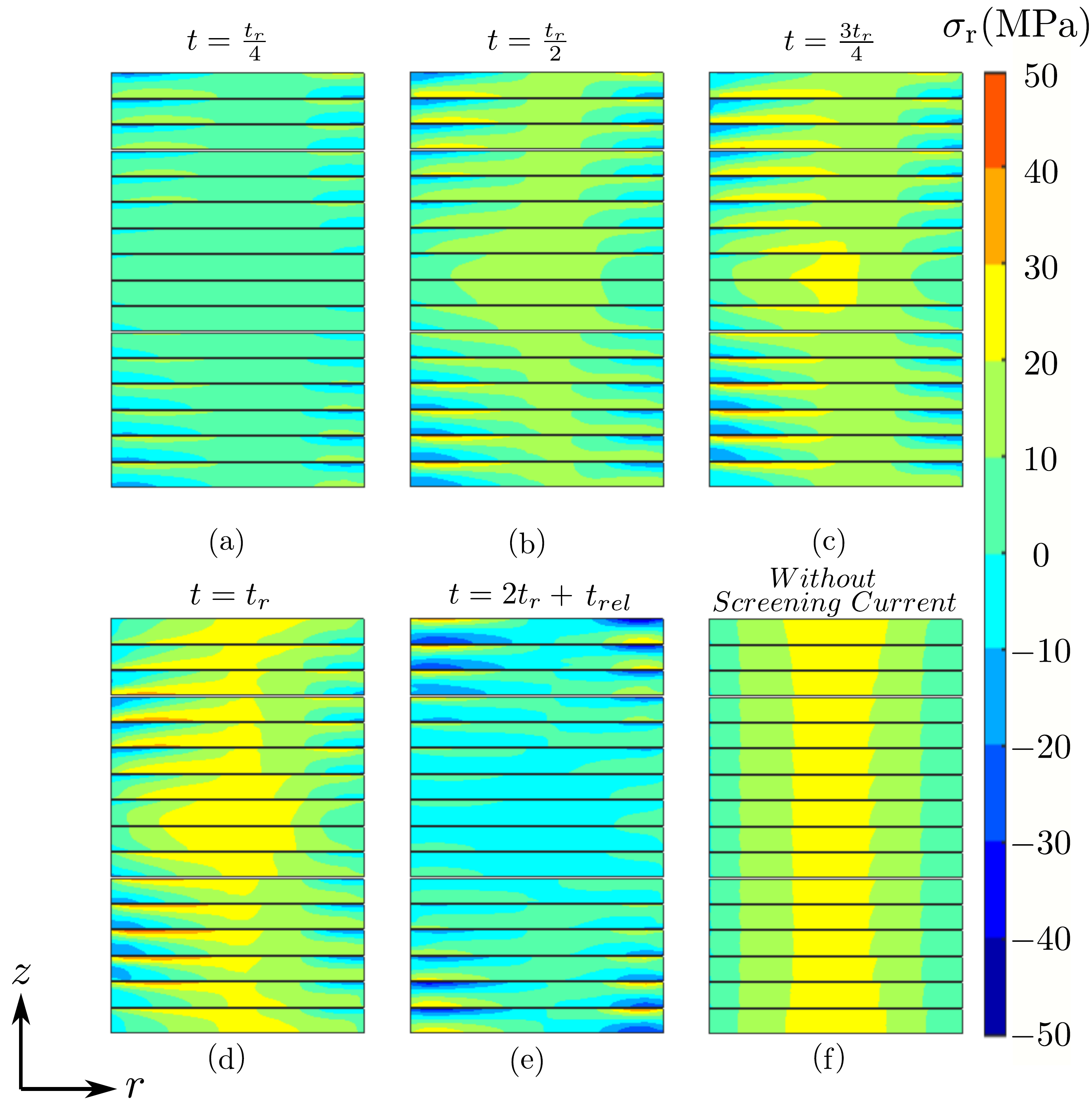}
     \caption{Evolution of Radial Stress $\sigma_{r}$ in a cycle for the whole magnet with screening current induced field. (a) to (c) Charging, (d) fully charged state, (e) discharged state and (f) without Screening current \R{at time $t=t_r$}.}
      \label{sigma_r}
\end{figure}

\begin{figure}
     \centering
    \includegraphics[scale=0.19]{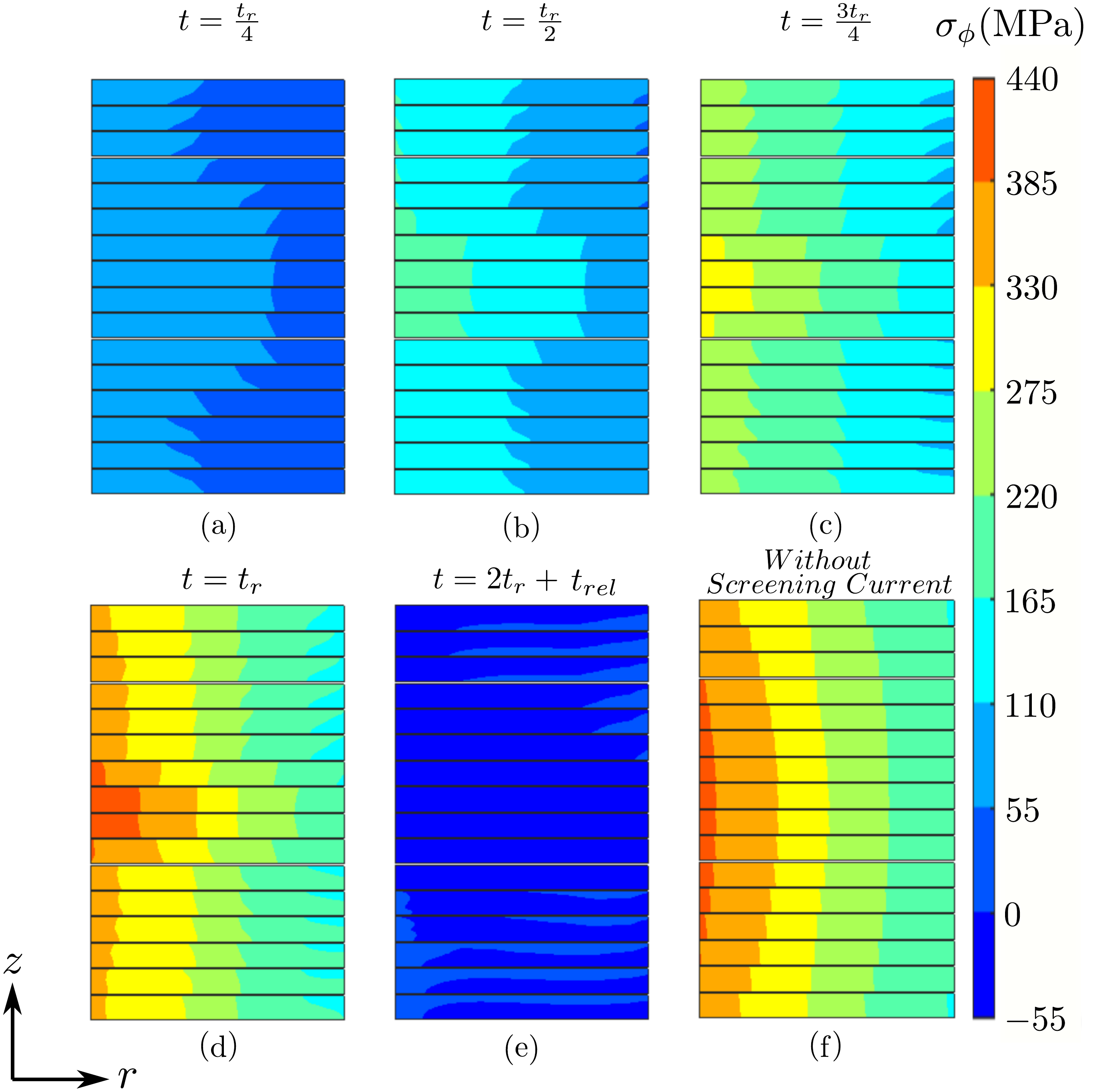}
     \caption{Variation of circumferential stress $\sigma_{\phi}$ in a cycle for the whole magnet with screening current induced field. (a) to (c) Charging, (d) fully charged state, (e) discharged state and (f) without Screening current \R{at time $t=t_r$}.}
      \label{sigma_phi}
\end{figure}

The evolution of circumferential strain for the charging and discharging cycle is shown in figure \ref{phi_strain}.
In agreement with literature \cite{yan2021screening,takahashi2020hoop}, the circumferential strain is maximum among all the other strain and should be of concern. For the present analysis the maximum circumferential strain is at the inner radius of the central pancakes. The order of $\epsilon_{\phi}$ is around  $0.25\%$  which is in line with our assumption that critical current density is independent of deformation. 

The calculation, taking SCIF into account, does not highly change the results. However, assuming uniform current density results in more uniform $\epsilon_\phi$(see, figure \ref{phi_strain}(f)). After discharging, the circumferential strain is both tensile and compressive. This is due to the fact that this only depends on the radial displacement and the displacement is positive and negative (see figure \ref{displacment}(e)) with in the pancake.

We also calculate the axial strain (see, figure \ref{axial_strain}). Curiously, the axial strain does not vanish, in spite of the fact that we neglect axial Lorentz forces. While charging, the axial strain in each pancake is  compressive in most of its cross section (see figure \ref{axial_strain}(d)). Due to the screening currents and the forces that they create, the sign of $\epsilon_z$ is anti-symmetirc with respect to the center of the coil cross-section. This is due to to the symmetry of the Lorentz force caused by the screening currents. The maximum magnitude of axial strain is at fully charged state is $0.12$ $\%$ and compressive, which appears at the central pancakes. The axial strains are more non homogeneous at the fully discharged state, as shown in figure \ref{axial_strain}(e).  Again, when there is no screening current the magnitude of axial strain is lower, which is now always compressive(see figure \ref{axial_strain}(f)).

Figure \ref{shear_strain} shows the shear strain in all the pancakes of the magnet. The shear effect only appears in the cases where SCIF is considered. The top half of the winding experiences positive shear while at the bottom opposite to it. The magnitude of maximum shear strain is around $0.09$ $\%$.

Interestingly, the shear changes its nature while discharging: the upper half of the whole winding experiences negative shear, being opposite at the bottom half. In addition, the magnitude increases to $0.12$ $\%$ even at zero current. 

When there is no screening current there is negligible shear as current density, and hence forces, are more uniform, see figure \ref{shear_strain}(f).

\begin{figure}
     \centering
    \includegraphics[scale=0.19]{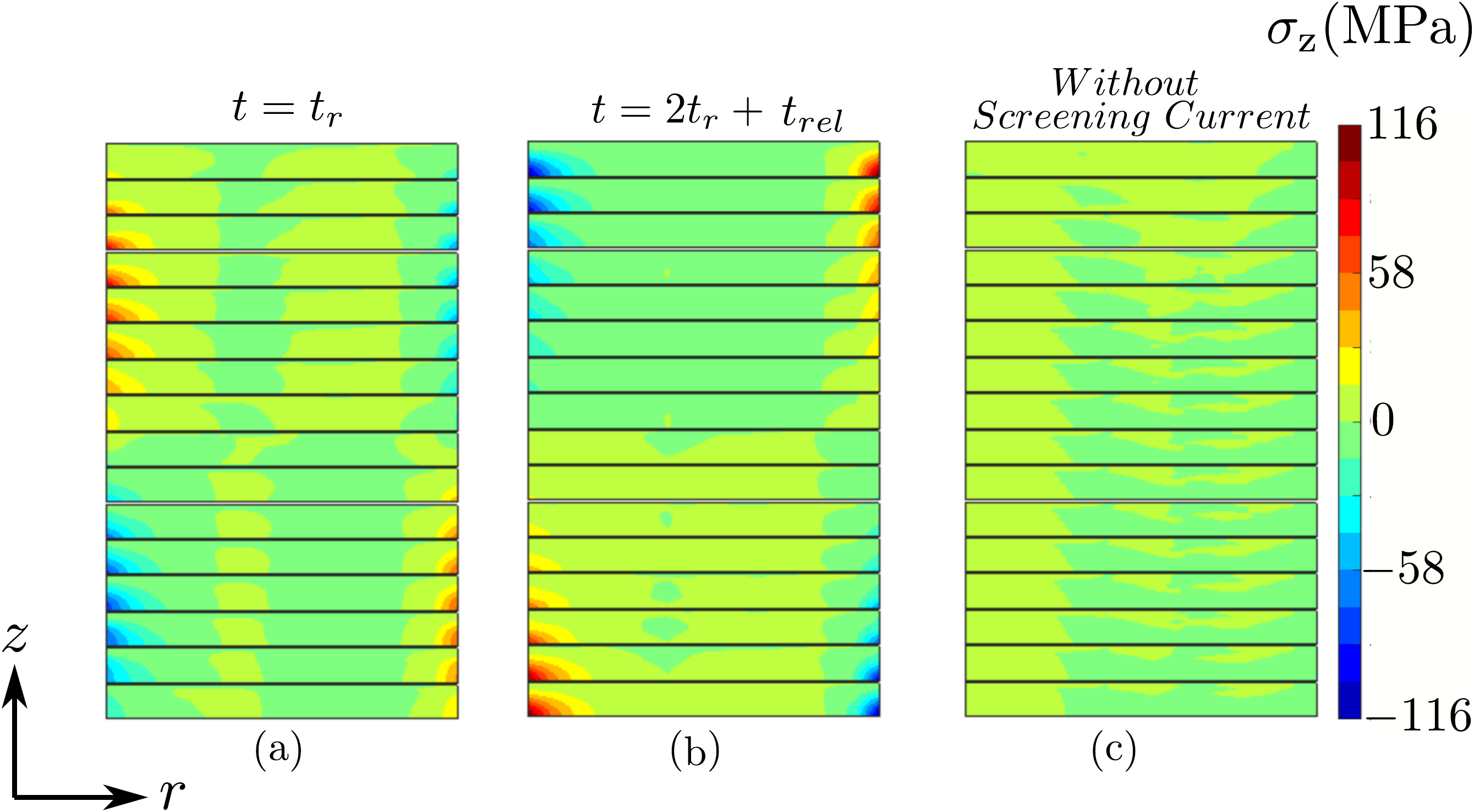}
    \caption{Variation of axial stress $\sigma_{z}$ in a cycle for the whole magnet with screening current induced field. (a) fully charged state, (b) discharged state and (c) without Screening current \R{at time $t=t_r$}.}
      \label{sigma_z}
\end{figure}

\begin{figure}
     \centering
    \includegraphics[scale=0.19]{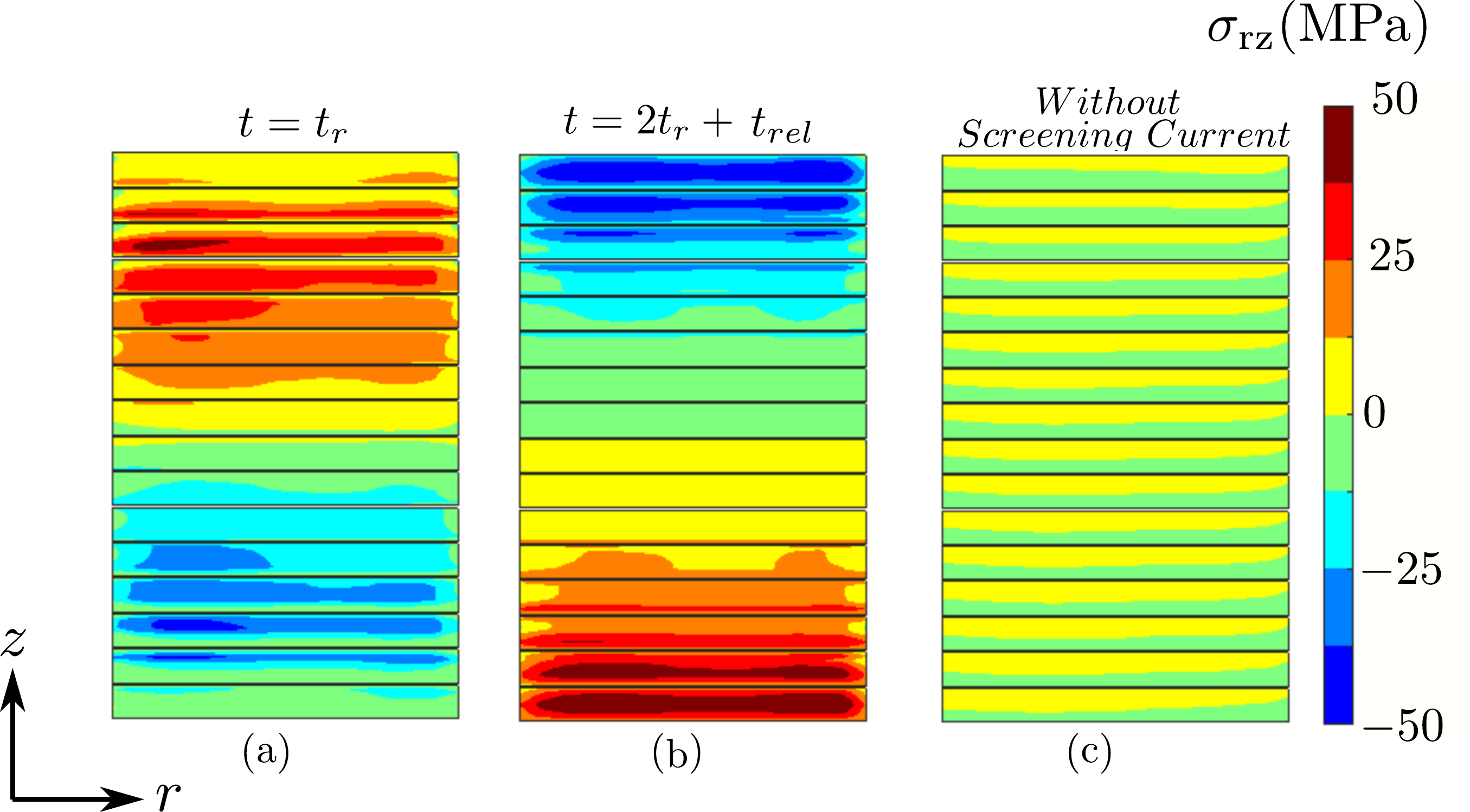}
    \caption{Variation of shear stress $\sigma_{rz}$ in a cycle for the whole magnet with screening current induced field. (a) fully charged state, (b) discharged state and (c) without Screening current \R{at time $t=t_r$}.}
      \label{sigma_rz}
\end{figure}

\subsubsection{Stress field in magnet:}
 Radial stresses for the magnet are plotted in figure \ref{sigma_r}. The radial stresses are increasing as the current is increasing. Due to non uniform screening currents each pancakes in the magnet is subjected to compressive and tensile stresses within the pancakes except in the central pancakes, where radial stresses are only tensile. This is due to the fact that central pancakes are subjected to more uniform Lorentz force density (see, figure \ref{radial_force}(d)). 
 
 When the magnet is fully charged, the maximum stress is $50$ MPa while the minimum is $-25$ $\mathrm{MPa}$. In fully discharged state the stresses are low in most region of magnet pancakes except around edges where it ranges from $50$ $\mathrm{MPa}$ to $-50$ MPa (see, figure \ref{sigma_r}(e)).  In addition, the radial stresses are opposite in sign close to the top and bottom of each pancake because the Lorentz forces (and current density) are also opposite in sign.

\begin{figure*}
     \centering
    \includegraphics[scale=0.6]{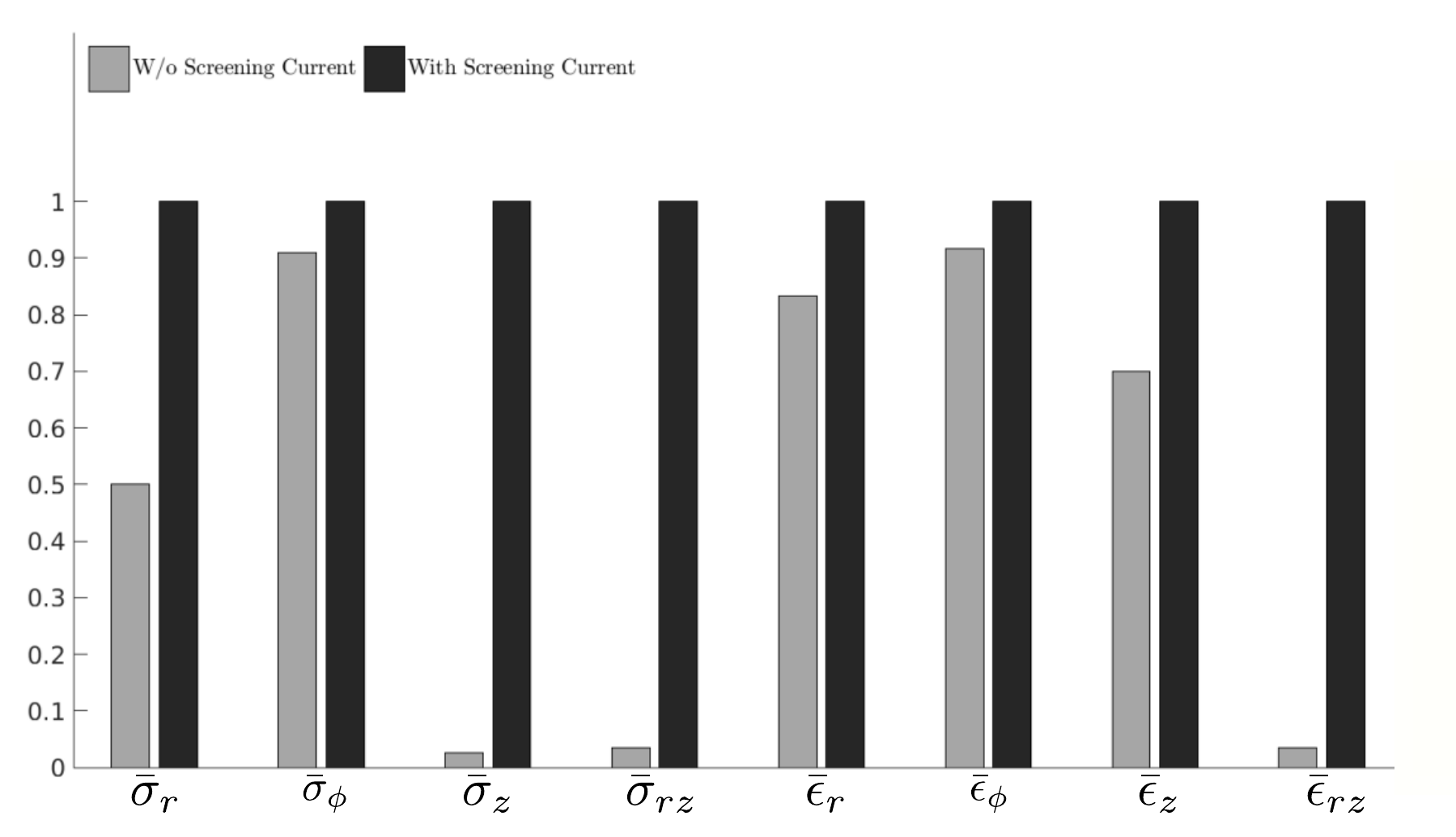}
     \caption{Comparison of mechanical quantities for magnet with and without screening current induced field.}
      \label{compare_mech}
\end{figure*}

 Figure\ref{sigma_r}(f) shows the distribution of radial stress for the case where there is no screening current. Now, the radial stresses are positive across the whole magnet. The maximum radial stresses are reduced to around $50$ $\%$ compared to the radial stresses in case with screening current for same input current and magnet geometry. Also, tensile radial stress is undesirable for design point of view, since all turns should be in physical contact with each other in a non-impregnated winding. Therefore, radial stress should be compressive at least in part of the tape width in order to enable physical contact between tapes. Otherwise, radial tensile stress could detach the tapes from each other in the radial direction. If detachment occurs, it will worsen the average thermal conductivity in the radial direction. In addition, the radial electric resistance will dramatically increase, which could be an issue for electro-thermal stability .

 In a soldered or impregnated winding, sufficiently large tensile stress could delaminate the superconducting layer, and hence permanently damage the winding. The critical delamination stress has been found to be as low as 25 MPa in certain tapes \cite{van2007delamination}.

 Figure\ref{sigma_phi} shows the time evolution of the of circumferential or hoop stress. The hoop stress is the highest  stress component. The maximum hoop stress is at the central pancakes and of magnitude $440$ $\mathrm{MPa}$, and tensile in nature. This value is much less than the critical value of tensile stress found in experiments for REBCO tapes \cite{barth2015electro}. Clearly, the model is able to capture the effect of screening current on hoop stresses(see, figure \ref{sigma_phi}(a-d)). In each pancake, the hoop stress is maximum at the inner radius and decreases with the radial coordinate. The top and bottom pancakes are subjected to lower stress due to screening currents. Interestingly, this trend is opposite from the radial stress.
 
  The nature of hoop stress after discharging the magnet is of concern, especially at the top and bottom pancakes (figure \ref{sigma_phi}(e)). These stresses are both tensile and compressive within each pancake leading to tensile and buckling effects at roughly the bottom and top halves of the pancake, respectively. Shunji \textit{et al} \cite{takahashi2020hoop} has shown such behaviour \R{and  suggested to bond the turns with epoxy to reduce this effect}. The advantage of model stem from the fact that it is able to capture such mechanics and hence it enables to take steps to diminish the forementioned effects. The hoop stress assuming no screening currents is more uniform (figure \ref{sigma_phi}(f)), though there is a small variation across the magnet.

We have also calculated the axial stresses in the magnet as shown in figure \ref{sigma_z}. For most of the magnet cross section these stresses are negligible. The axial stresses in these magnet are such that it tries to topple the pancake in clockwise direction in the pancakes at the upper half of the winding and vice versa for pancakes that are on the bottom half. This phenomenon is justified as each pancake is fixed at its bottom by roller boundaries, which impose no axial displacements at that boundary ($u_z=0$).. The maximum magnitude of axial stress is $110$ $\mathrm{MPa}$.

The shear stress is plotted in figure \ref{sigma_rz} which evolves similarly to shear strain during charging and discharging. Again, when there is no screening current the shear stress is negligible. The maximum amount of shear stress at fully charged state (see figure \ref{sigma_rz}(d)) is  $50$ $\mathrm{MPa}$. At fully discharged state, the magnet is also subjected to shear stress due to screening current induced field from the superconductor. 

The effect of shear stress on REBCO superconducting tapes has not been well studied, but we could expect that tapes will be able to withstand much lower shear than hoop stresses. This is due to the brittle nature of the superconductor and the buffer layers.

\begin{figure}
     \centering
    \includegraphics[scale=0.19]{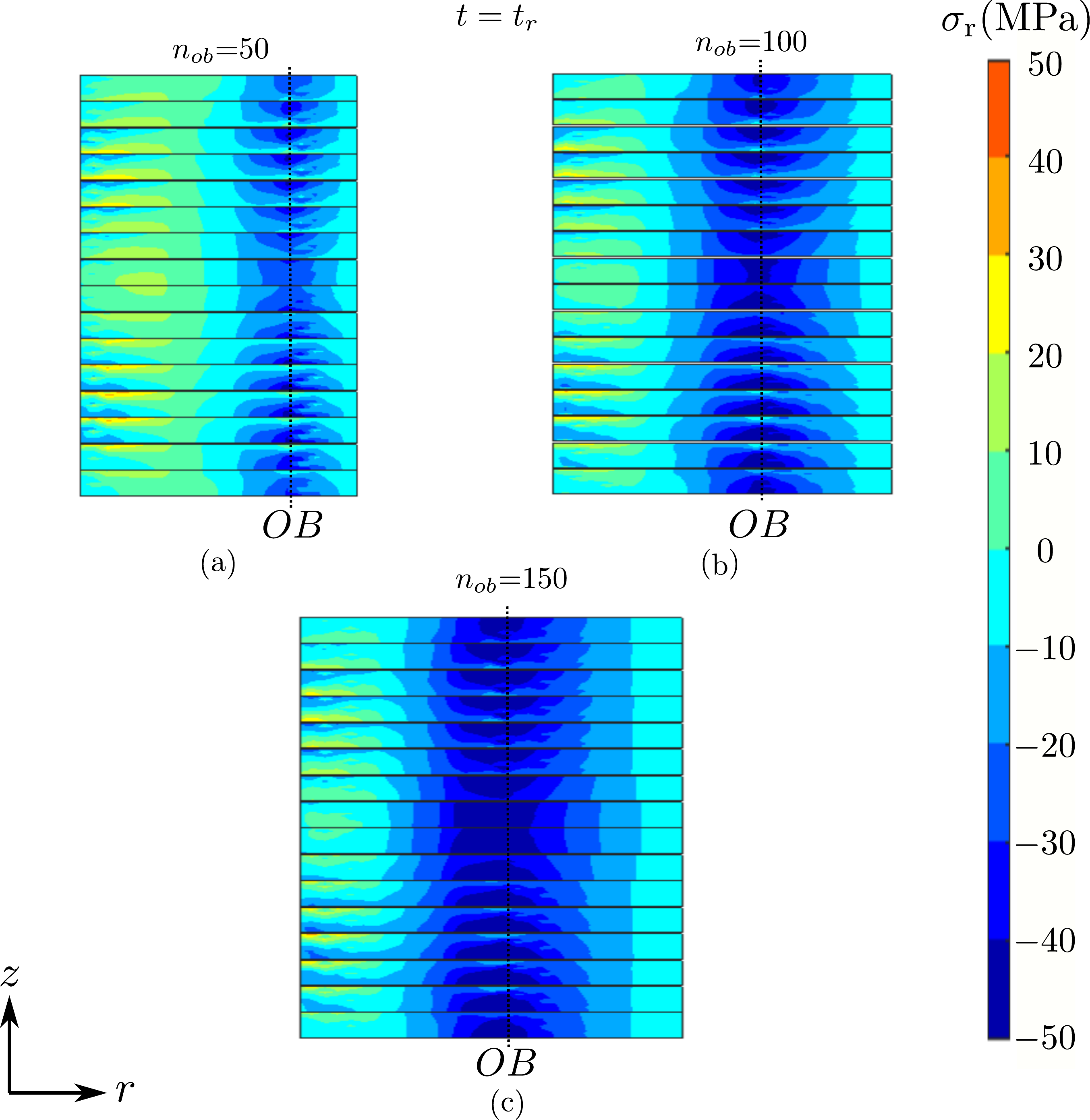}
    \caption{Variation of radial stress in the magnet subjected to screening current induced field when number of overbanding turns are (a) $n_{ob}=50$, (b) $n_{ob}=100$ and (c) $n_{ob}=150$. \R{The dotted vertical line shows the boundary between REBCO coil and overbanding coil}.}
      \label{sigma_r_sc_ob}
\end{figure}

\begin{figure}
     \centering
    \includegraphics[scale=0.19]{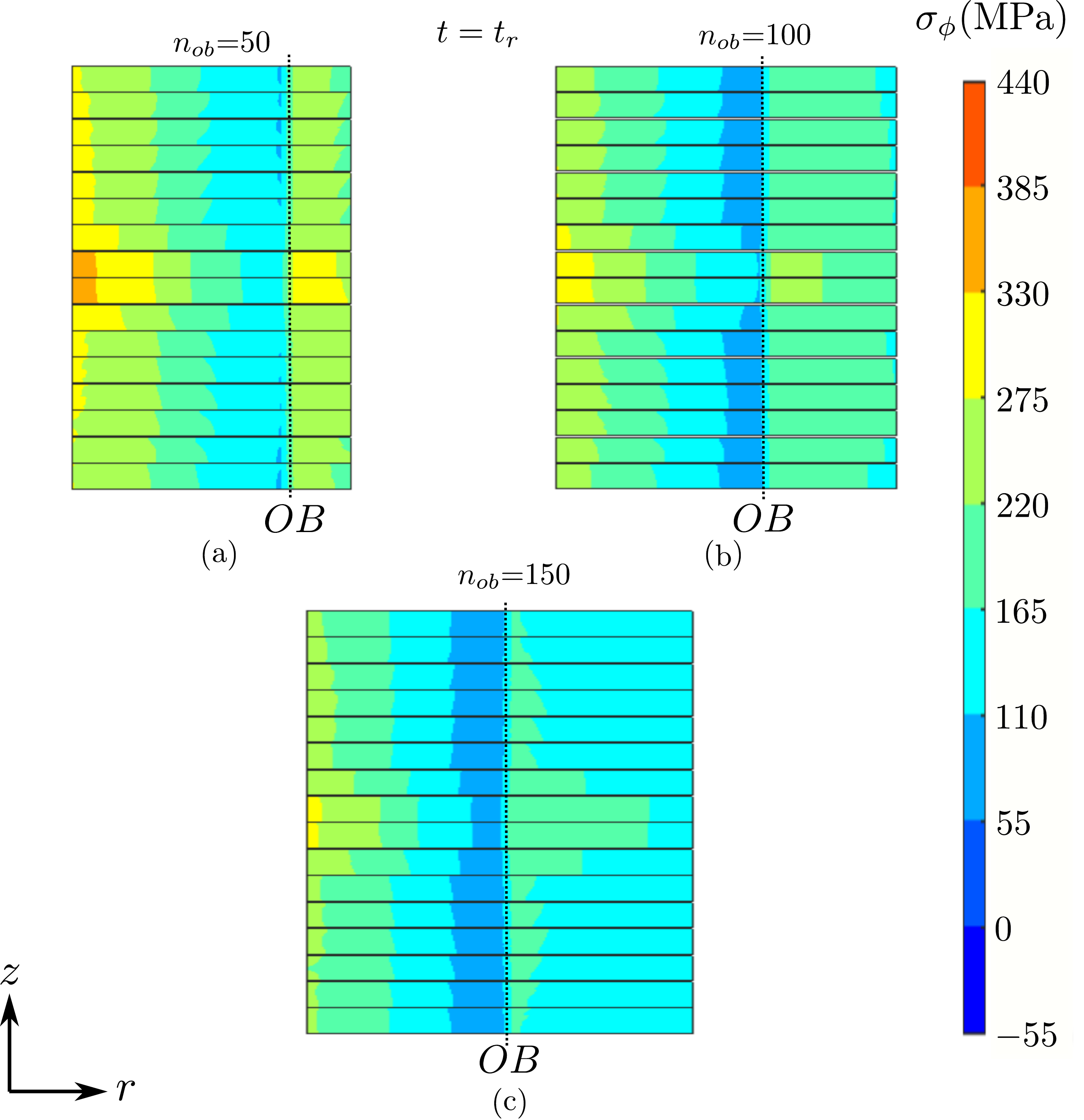}
     \caption{Variation of circumferential stress in the magnet subjected to screening current induced field when number of overbanding turns are (a) $n_{ob}=50$, (b) $n_{ob}=100$ and (c) $n_{ob}=150$.\R{The dotted vertical line shows the boundary between REBCO coil and overbanding coil}.}
          \label{sigma_phi_sc_ob}
\end{figure}

\begin{figure}
     \centering
    \includegraphics[scale=0.19]{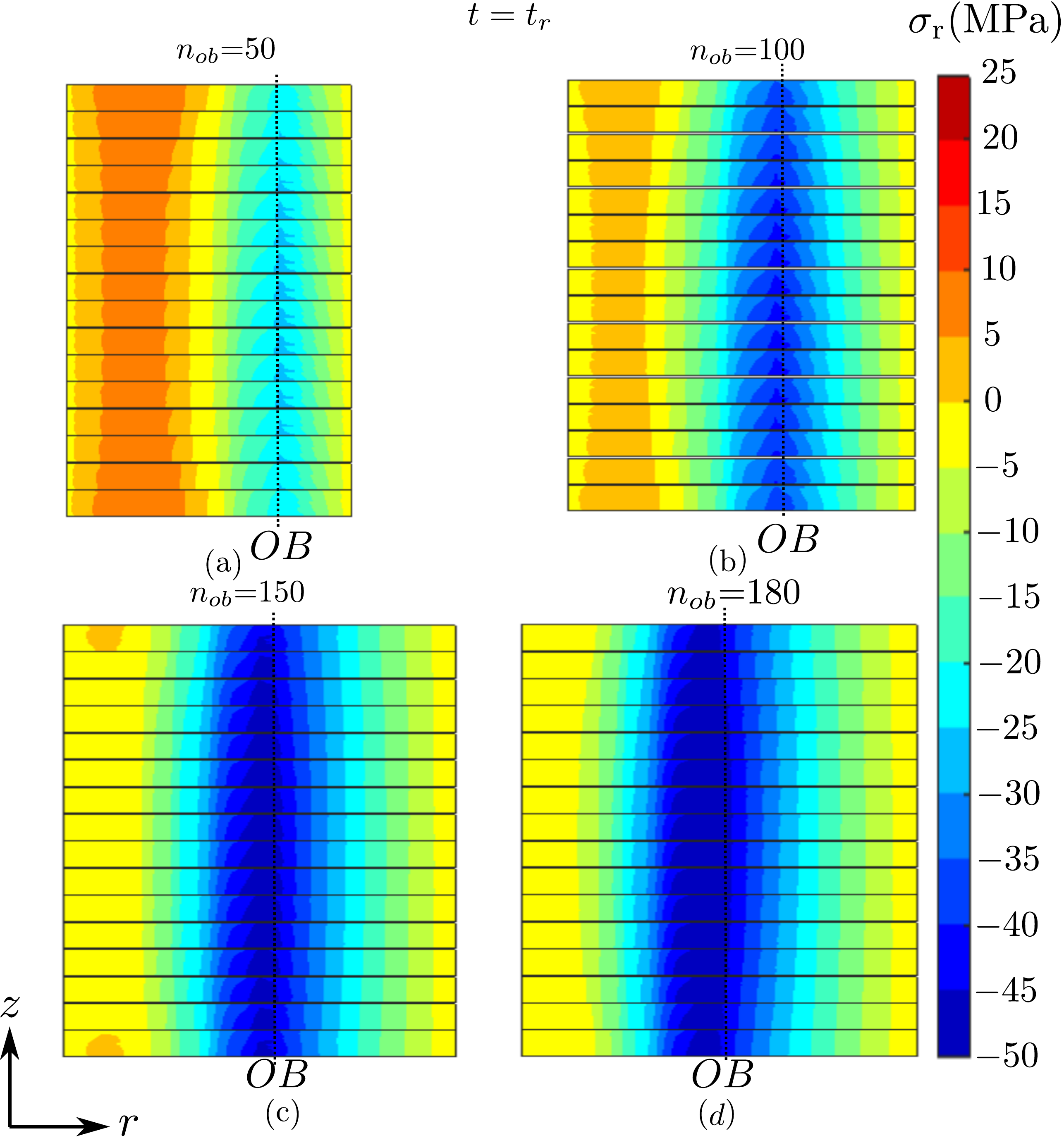}
     \caption{Variation of radial stress in the magnet subjected to no screening current induced field when number of overbanding turns are (a) $n_{ob}=50$, (b) $n_{ob}=100$, (c) $n_{ob}=150$ and (c) $n_{ob}=180$. \R{The dotted vertical line shows the boundary between REBCO coil and overbanding coil}.}
      \label{sigma_r_nosc_ob}
\end{figure}

\begin{figure*}
     \centering
    \includegraphics[scale=0.5]{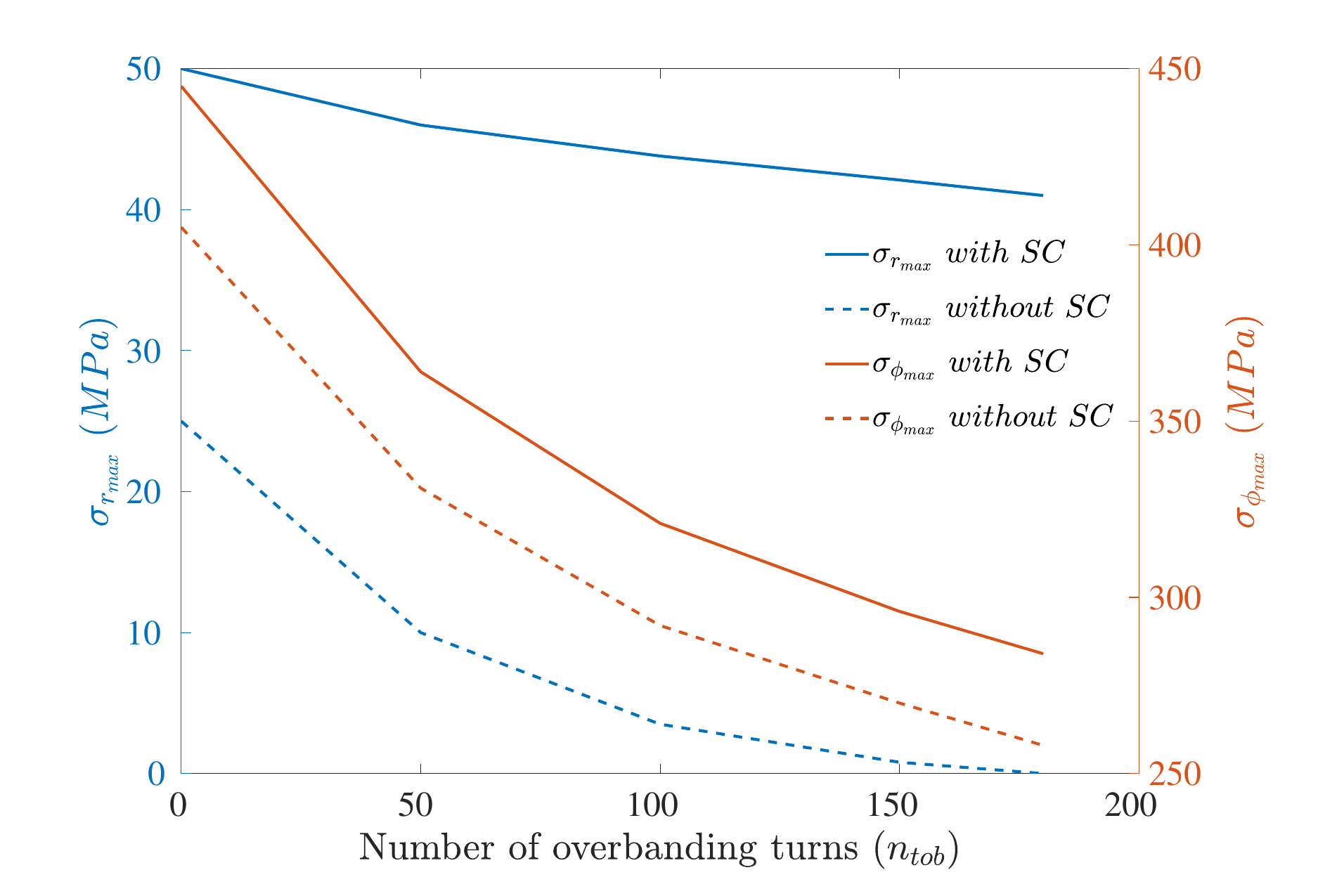}
     \caption{Variation of peak radial as well as circumferential stress with number of overbanding turns ($n_{ob}$). The solid lines represents SCIF while dotted lines correspond to no SCIF.}
      \label{compare_overbanding}
\end{figure*}

\subsubsection{Relevance of screening currents}:
Next, we compare the peak quantities with and without screening currents. Figure \ref{compare_mech} shows normalized stresses $\bar{\sigma}=\sigma_{max}/\sigma_{{sc}_{max}}$ and normalized strains $\bar{\epsilon}=\epsilon_{max}/\epsilon_{{sc}_{max}}$ for all components. The effect of screening current in amplifying the peak values is clearly demonstrated. Clearly, the screening currents increase all components of stress and strain.The strain increase is the most pronounced for the shear component, which is negligible for no screening currents but might be critical when screening currents are taken into account. The radial stress also increases substantially, by a factor 2, but the hoop stress increases by only around 10 \%. , which is similar to the safety margin regarding the operating current. Then screening currents play an essential role for the shear and radial stress, but they are only relevant for the hoop stress when this component is close to the critical value.

\subsection{Overbanding}
This section discuss the effect of over banding as shown in figure \ref{OB_schematic}. We consider a stiff over banding of Tungsten carbide, whose young modulus is $650 MPa$.  The thickness of overbanding layer is the same as the REBCO tape for the numerical analysis. The winding is then subjected to similar input current and external fields as in section \ref{results_EM}.  It is  shown that using overbanding reduce the tensile stresses. 

There different overbanding turns are winded on the tape and the mechanics is analyzed for the magnet. We have only shown here the effect of overbanding winding for magnet coil on radial and circumferntial stresses.

First, we analyze the radial stress taking screening currents into account. Figure  \ref{sigma_phi_sc_ob} shows that there is always some region with tensile stress, even for large number of turns of overbanding. However, for 100 or more turns of overbanding there is always a region within the pancake when the radial stresses are compressive for every turn of each pancake, which indicates that they are in contact with their neighbour turns. This region increases with increase in the number of overbanding turns as shown, in figure \ref{sigma_r_sc_ob}. In addition, the maximum stresses are reduced from  $50$ $\mathrm{MPa}$ for no overbanding to  $20$ $\mathrm{MPa}$ for $n_{ob}=150$. At the boundary between the two materials the stresses are highly compressive which is reasonable because the overbanding layer acts nearly to a rigid boundary against coil expansion.

Next, we analyse the circumferential stress, which shows that this stress decreases with the number of overbanding turns (see, figure \ref{sigma_phi_sc_ob}). 

If we neglect the screening currents, the radial stresses are relatively low which are tensile in a large portion of the superconducting winding for 50 and 100 turns of overbanding, only at the top and bottom pancakes for $n_{ob}=150$. The radial stress is compressive for the whole superconducting cross-section for large enough turns of overbanding (180 in our case, see, figure \ref{sigma_r_nosc_ob}).

Finally, figure \ref{compare_overbanding} shows the dependence of the paeak radial and circumferential stress with the number of overbanding turns. Both peak radial and circumferential stresses are decrease with the number of overbanding turns. However, overbanding causes a higher decrease in the maximum circumferential stress than in the radial component.

\section{Conclusions}
 Superconducting high field magnets are subjected to strong Lorentz forces that could damage the superconductor and other structural materials. Then, the design of high-field magnets require mechanical modelling in order to avoid damage during operation. We use this method to analyze a REBCO insert in a 32 T fully superconducting magnet.  We have included the effect of screening current induced field to calculate the mechanical properties. These magnets have to function safely when subjected to high current density in order to generate highest possible magnetic field. To ensure and checking this, a fast and accurate tool is developed. Following are the salient point of the work:

\begin{enumerate}
    \item Bulk approximation to calculate elastic properties leads to good agreement with uniaxial experiments. 

    \item The top and the bottom of the pancakes are more susceptible to screening currents,which cause non uniform current density and the increased strains and stresses.

    \item The radial strains are compressive and maximum at the inner radius.

    \item The highest component among all the strains and stresses are circumferential. Both increase with the radial coordinate.

    \item The shear stresses and shear strains arise due to SCIF. Although much lower than the circumferential stress, they are a matter of concern for delamination in REBCO tapes. These mechanical quantities appear as negligible when screening currents effect are not taken into account. \R{The value of shear stress is somewhat high due to the limitation of model taking not into account of mechanics in between layers. A proper modification is needful for more accurate prediction.}

    \item After discharging the REBCO insert, winding is subjected to stress, which are caused by SCIF. The circumferential stress is of concern as it may lead to buckling.

   \R{ \item The value of radial, axial and hoop stresses are with in permissible limit with use of proper overbanding material. Also, stresses will decrease and more realistic if we include further stress modification.}

    \item The use of stiff overbanding material reduces the radial and circumferential stresses which protect the magnet from damage.

\end{enumerate}

    \R{For more realistic modelling, further stress modification and thermal aspect should be included in modelling. Adding additional stress due to winding pre-tension will further reduce the stresses, particularly radial stresses. Also, there is evidence of delamination of REBCO layer which lead to damage of coils. There is also need to include this factor while designing these HIGH field magnet. We are also hopeful that with advancement in manufacturing technique for REBCO tapes there adhesive strength will increase. }

\section*{Acknowledgement}
A.S. acknowledges  simulating discussion with Dr. Anang Dadhich at various occasion during the work. E.P. and A.S. acknowledge Oxford Instruments for providing details on the cross-section of the LTS outsert. This project has received funding from the European Union's Horizon 2020 research and innovation programme under grant agreement No 951714 (superEMFL), and the Slovak Republic from projects APVV-19-0536 and VEGA 2/0098/24. Any dissemination of results reflects only the author's view and the European Commission is not responsible for any use that may be made of the information it contains.
\section*{References}


\end{document}